\documentclass[preprint2,numberedappendix]{emulateapj-rtx4}
\usepackage{graphicx,times,bm,url}
\graphicspath{{./fig/}{./png/}}

\newcommand{\EQ}{\begin{equation}}
\newcommand{\EN}{\end{equation}}
\newcommand{\EQA}{\begin{eqnarray}}
\newcommand{\ENA}{\end{eqnarray}}

\newcommand{\EEq}[1]{Equation~(\ref{#1})}
\newcommand{\Eq}[1]{Eq.~(\ref{#1})}
\newcommand{\Eqs}[2]{Eqs.~(\ref{#1}) and~(\ref{#2})}
\newcommand{\Eqss}[2]{Eqs.~(\ref{#1})--(\ref{#2})}

\newcommand{\App}[1]{Appendix~\ref{#1}}
\newcommand{\Sec}[1]{Sect.~\ref{#1}}

\newcommand{\Fig}[1]{Fig.~\ref{#1}}
\newcommand{\FFig}[1]{Figure~\ref{#1}}
\newcommand{\Figs}[2]{Figs.~\ref{#1} and \ref{#2}}

\newcommand{\Tab}[1]{Table~\ref{#1}}

\newcommand{\bra}[1]{\langle #1\rangle}

\newcommand{\meanrho}{\overline{\rho}}
\newcommand{\meanPhi}{\overline{\Phi}}
{}
{}
\newcommand{\meanFFFF}{\overline{\mbox{\boldmath ${\cal F}$}}{}}{}

\newcommand{\meanSSSS}{\overline{\mbox{\boldmath ${\mathsf S}$}} {}}

{}
{}
\newcommand{\meanEMF}{\overline{\mbox{\boldmath ${\cal E}$}}{}}{}
{}
{}
{}
{}
\newcommand{\meanAA}{\overline{\mbox{\boldmath $A$}}{}}{}
\newcommand{\meanBB}{\overline{\mbox{\boldmath $B$}}{}}{}
{}
{}
{}
{}
{}
{}
{}
\newcommand{\meanJJ}{\overline{\mbox{\boldmath $J$}}{}}{}
\newcommand{\meanUU}{\overline{\bm{U}}}

{}
{}
{}

\newcommand{\meanB}{\overline{B}}

\newcommand{\meanU}{\overline{U}}

\newcommand{\meanJ}{\overline{J}}

{}

{}
{}

\newcommand{\ghat}{\hat{g}}
\newcommand{\gghat}{\hat{\bm{g}}}
\newcommand{\uu}{\mbox{\boldmath $u$} {}}
\newcommand{\UU}{\mbox{\boldmath $U$} {}}

\newcommand{\bb}{\mbox{\boldmath $b$} {}}
\newcommand{\BB}{\mbox{\boldmath $B$} {}}

\newcommand{\JJ}{\mbox{\boldmath $J$} {}}

\newcommand{\AAA}{\mbox{\boldmath $A$} {}}

\newcommand{\ff}{\mbox{\boldmath $f$} {}}

\newcommand{\grav}{\mbox{\boldmath $g$} {}}
\newcommand{\nab}{\mbox{\boldmath $\nabla$} {}}

\newcommand{\ggamma}{\mbox{\boldmath $\gamma$} {}}

\newcommand{\SSSS}{\mbox{\boldmath ${\sf S}$} {}}

\newcommand{\ii}{{\rm i}}

\newcommand{\DD}{{\rm D} {}}

\newcommand{\dd}{{\rm d} {}}

\newcommand{\const}{{\rm const}  {}}

\def\Pm{\mbox{\rm Pr}_M}
\def\Rm{\mbox{\rm Re}_M}

\def\Rey{\mbox{\rm Re}}

\def\qpz{q_{p0}}
\def\qsz{q_{s0}}
\def\qp{q_{p}}
\def\qs{q_{s}}
\def\qe{q_{g}}
\def\qg{q_{g}}

\def\cs{c_{s}}
\def\Pturb{P_{\rm turb}}

\def\ellf{\ell_{f}}

\def\kf{k_{f}}

\def\urms{u_{\rm rms}}
\def\urmsz{u_{\rm rms0}}

\def\nut{\nu_{t}}
\def\etat{\eta_{t}}
\def\etatz{\eta_{t0}}

\def\betap{\beta_{p}}
\def\betamin{\beta_{\rm min}}
\def\betarms{\beta_{\rm rms}}
\def\betastar{\beta_{\star}}

\def\Beq{B_{\rm eq}}
\def\Beqz{B_{\rm eq0}}
\def\Peff{{\cal P}_{\rm eff}}
\def\Pmin{{\cal P}_{\rm min}}
\def\Bmin{B_{\rm min}}

\def\half{{\textstyle{1\over2}}}

\def\onethird{{\textstyle{1\over3}}}
\def\onesixth{{\textstyle{1\over6}}}

\def\kd{k_{d}}
\def\kf{k_{f}}

\newcommand{\yapj}[3]{ #1, {ApJ,} {#2}, #3}

\newcommand{\yapjl}[3]{ #1, {ApJ,} {#2}, #3}

\newcommand{\yan}[3]{ #1, {Astron.\ Nachr.,} {#2}, #3}

\newcommand{\yana}[3]{ #1, {A\&A,} {#2}, #3}

\newcommand{\ygafd}[3]{ #1, {Geophys.\ Astrophys.\ Fluid Dyn.,} {#2}, #3}

\newcommand{\yjfm}[3]{ #1, {J.\ Fluid Mech.,} {#2}, #3}

\newcommand{\ypf}[3]{ #1, {Phys.\ Fluids,} {#2}, #3}
\newcommand{\ypfb}[3]{ #1, {Phys.\ Fluids B,} {#2}, #3}

\newcommand{\ysovl}[3]{ #1, {Sov.\ Astron.\ Lett.,} {#2}, #3}
\newcommand{\yjetp}[3]{ #1, {Sov.\ Phys.\ JETP,} {#2}, #3}

\newcommand{\yanf}[3]{ #1, {Ann. Rev. Fluid Mech.,} {#2}, #3}
\newcommand{\yprs}[3]{ #1, {Proc.\ Roy.\ Soc.\ Lond.,} {#2}, #3}

\newcommand{\yprl}[3]{ #1, {Phys.\ Rev.\ Lett.,} {#2}, #3}

\newcommand{\ymn}[3]{ #1, {MNRAS,} {#2}, #3}
\newcommand{\ynat}[3]{ #1, {Nature,} {#2}, #3}
\newcommand{\yrpp}[3]{ #1, {Rep.\ Progr.\ Phys.,} {#2}, #3}
\newcommand{\yptrsa}[3]{ #1, {Phil. Trans. Roy. Soc. London A,} {#2}, #3}

\newcommand{\ysph}[3]{ #1, {Solar Phys.,} {#2}, #3}

\newcommand{\ypre}[3]{ #1, {Phys.\ Rev.\ E,} {#2}, #3}

\newcommand{\yjour}[4]{ #1, {#2}, {#3}, #4}

\newcommand{\ybook}[3]{ #1, {#2} (#3)}

\begin{document}
\title{The negative effective magnetic pressure in stratified forced turbulence}

\author{Axel Brandenburg, Koen Kemel}
\affil{
NORDITA, AlbaNova University Center, Roslagstullsbacken 23,
SE-10691 Stockholm, Sweden; \\
Department of Astronomy, AlbaNova University Center,
Stockholm University, SE-10691 Stockholm, Sweden
}
\author{Nathan Kleeorin, Igor Rogachevskii}
\affil{
Department of Mechanical
Engineering, Ben-Gurion University of the Negev, POB 653,
Beer-Sheva 84105, Israel;\\
NORDITA, AlbaNova University Center, Roslagstullsbacken 23,
SE-10691 Stockholm, Sweden
}

\date{~$ $Revision: 1.269 $ $}

\begin{abstract}
To understand the basic mechanism of the formation of magnetic flux
concentrations, we determine by direct numerical simulations the
turbulence contributions to the mean magnetic pressure in a strongly
stratified isothermal layer with large plasma beta, where a weak
uniform horizontal mean magnetic field is applied. The negative
contribution of turbulence to the effective mean magnetic pressure
is determined for strongly stratified forced turbulence over a range
of values of magnetic Reynolds and Prandtl numbers. Small-scale dynamo
action is shown to reduce the negative effect of turbulence on the
effective mean magnetic pressure. However, the turbulence
coefficients describing the negative effective magnetic pressure
phenomenon are found to be converged for magnetic Reynolds numbers
between 60 and 600, which is the largest value considered here. In
all these models the {\it turbulent intensity}
is arranged to be nearly independent of
height, so the kinetic energy density decreases with height due to
the decrease in density. In a second series of numerical
experiments, the turbulent intensity increases with height such that
the {\it turbulent kinetic energy density} is nearly independent of
height. Turbulent magnetic diffusivity and turbulent pumping
velocity are determined with the test-field method for both cases.
The vertical profile of the turbulent magnetic diffusivity is found
to agree with what is expected based on simple mixing length
expressions. Turbulent pumping is shown to be down the gradient of
turbulent magnetic diffusivity, but it is twice as large as
expected. Corresponding numerical mean-field models are used to show
that a large-scale instability can occur in both cases, provided the
degree of scale separation is large enough and hence the turbulent
magnetic diffusivity small enough.
\end{abstract}

\section{Introduction}

In a stratified layer, magnetic fields do not normally stay in equilibrium
but tend to become buoyantly unstable
\citep[e.g.][]{New61,Par66,Par79a,Gil70a,Gil70b,HP88,Catt88,Wis00,Isobe05,Kers07},
see also reviews by \cite{Hugh07} and \cite{Tob07}.
This mechanism related to magnetic buoyancy,
is generally invoked in order to understand magnetic
flux emergence at the solar surface \citep[e.g.][]{HAGM09}.
The mechanism does not explicitly rely upon the existence of turbulence,
except that the origin of the Sun's magnetic field is
generally believed to be turbulent in nature;
see \cite{Solanki} for a recent review.

Turbulent dynamos work under a variety of
circumstances and are able to produce weakly nonuniform
large-scale magnetic fields
(see Brandenburg \& Subramanian 2005 for a review).
At first glance this generation process is counter-intuitive, because
it works against the well-known concept of turbulent mixing
\citep{Tay21,Pra25}.
However, it is now well established that turbulence can also
cause non-diffusive effects.
In addition to the well-known $\alpha$ effect that is generally
believed to be responsible for the Sun's large-scale field
\citep[][]{Mof78,Par79b,KR80},
there is also the $\Lambda$ effect that is responsible for driving
the differential rotation of the Sun \citep{Rue80,Rue89,RH04}.
Yet another import effect is turbulent pumping or $\gamma$ effect
\citep{Rad69}, which corresponds to the advection of mean magnetic
field that is not associated with any material motion.
The $\gamma$ effect appears, for example,
in nonuniform turbulence and transports
mean magnetic field down the gradient of turbulent intensity,
which is usually downward in turbulent convection.
However, this effect can also be modified by the mean magnetic
field itself \citep{Kitch94,RK06}, which can then correspond to a
mean-field buoyancy effect.

When invoking the concept of magnetic buoyancy, one must ask what
the effect of turbulence is in this context. The turbulent pressure
associated with the convective fluid motions and magnetic
fluctuations is certainly not negligible and reacts sensitively to
changes in the background magnetic field. The main reason for this
is that the kinetic energy density in isotropic turbulence
contributes to the total turbulent dynamic pressure twice as much as
turbulent magnetic energy density \citep[][hereafter referred to as
RK07]{KRR89,KRR90,RK07}:
\EQ
\Pturb={\textstyle{1\over3}}\overline{\rho \uu^2}
+{\textstyle{1\over 6}}\overline{\bb^2}/\mu_0.
\EN
Here, $\Pturb$ is
the total turbulent dynamic pressure caused by velocity and magnetic
fluctuations, $\uu$ and $\bb$, respectively, $\mu_0$ is the vacuum
permeability, $\rho$ is the fluid density, and overbars indicate
ensemble averaging.
On the other hand, any rise in local turbulent
magnetic energy density must be accompanied by an equal and opposite
change of turbulent kinetic energy density in order to obey
approximate energy conservation, i.e.\
\EQ
\half\overline{\rho\uu^2}+\half\overline{\bb^2}/\mu_0\equiv E_{\rm
tot}\approx\const.
\EN
Direct numerical simulations in open systems
with boundaries \citep[][hereafter referred to as BKR]{BKR10} show
that when the mean magnetic field $\meanB$ is much smaller than the
equipartition field strength, $\Beq$, the total energy is conserved, while
when $\meanB \leq \Beq$, $E_{\rm tot}$ decreases slightly with increasing
mean field, and it varies certainly less than either $\overline{\rho\uu^2}$
or $\overline{\bb^2}$; see Fig.~1 of BKR.
This clearly implies that,
upon generation of magnetic fluctuations, the total turbulent
dynamic pressure shows a reversed (destabilizing) feedback
\citep{KRR90}, i.e.\
\EQ
\Pturb=-\onesixth\overline{\bb^2}/\mu_0+2E_{\rm tot}/3,
\label{phenomenology}
\EN
so both an increase of $\overline{\bb^2}$,
as well as an increase of the imposed field, which decreases $E_{\rm
tot}$, tend to lower the value of $\Pturb$.
For strongly anisotropic
turbulence, \Eq{phenomenology} is also valid except for the change
of the $1/6$ factor into $1/2$ (RK07). This phenomenology was
supported by analytical studies using the spectral $\tau$ relaxation
approximation \citep{KRR90} and the renormalization approach
\citep{KR94} and led to the realization that the effective mean
magnetic pressure force (the sum of turbulent and non-turbulent
contributions) is reduced and can be reversed for certain mean
magnetic field strengths. Under certain conditions (e.g.\ strong
density stratification), this can cause a magnetic buoyancy
instability via perturbations of a uniform mean magnetic field in
stratified turbulence \citep{KMR93,KMR96}. Later, when considering
the effect of turbulent convection on the mean Lorentz force, RK07
suggested that magnetic flux concentrations in the Sun such as
active regions and even sunspots might be formed by this reversed
feedback effect.

Most of the numerical simulations on magnetic flux emergence
\citep[e.g.][]{Stein01,Schus06,Mart08,Remp09} have been done using initial
conditions with an already existing strongly inhomogeneous
large-scale magnetic field.
Recent simulations by BKR and \cite{KKWM10,Kapy11} study
the formation of large-scale magnetic structures from an
initially uniform large-scale magnetic field. In particular,
Large-Eddy Simulations of solar magneto-convection by \cite{KKWM10}
give indications that the spontaneous formation of long-lived
magnetic flux concentrations from an initial vertical
uniform magnetic field might be possible, although the
underlying mechanism in their simulations still remains to be clarified.
A similar type of magnetic flux concentration
with vertical imposed field has been seen in
convection simulations at large aspect ratios by \cite{Tao98}, which
show a segregation into magnetized and weakly magnetized regions.
One of the differences compared with BKR is the vertical orientation
of the imposed magnetic field in turbulent convection.
In forced and convection-driven turbulence simulations of BKR and
\cite{Kapy11}, respectively, the imposed magnetic field was a horizontal one.
Other possibilities for causing flux concentrations include turbulent
thermal collapse, whereby the magnetic field suppresses the convective
energy flux, leading to local cooling, and thus to contraction and further
enhancement of magnetic flux \citep{KM00}.
By considering an isothermal equation of state with isothermal stratification,
we will exclude this possibility in our present work, allowing thus a
more definitive identification of the effect of density-stratified
turbulence on the mean Lorentz force.

Meanwhile, direct numerical simulations of forced turbulence
with an imposed horizontal magnetic field have demonstrated conclusively
that in a simulation with an isothermal equation of state
and an isothermal density stratification, spontaneous
formation of magnetic structures \citep{BKKMR11,KBKMR}
does indeed occur.
Those simulations used a scale separation ratio of 1:15 and 1:30,
i.e., the computational domain must be big enough to encompass at least
15 (or even 30) turbulent eddies in one coordinate direction.
It does then become computationally expensive to achieve large Reynolds
numbers, because their value is based on the size of the energy-carrying
turbulent eddies rather than the size of the computational domain.
In the present paper we restrict ourselves to a scale separation ratio
of 1:5 and are thereby able to demonstrate convergence of the turbulence
coefficients describing the negative effective magnetic pressure
phenomenon for magnetic Reynolds numbers between 60 and 600.

The basic phenomenon of magnetic flux concentration
by the effect of turbulence on the mean Lorentz force
has been studied by BKR based on numerical solutions
of the {\it mean-field} momentum and induction equations.
They demonstrated the existence of
a linear instability for sufficiently strong stratification.
This instability was followed by nonlinear saturation
at near-equipartition strengths.
Using direct numerical simulations (DNS) of forced turbulence,
BKR also verified the
validity of the phenomenology highlighted by \Eq{phenomenology}.
However, their DNS ignored the effects of stratification
which would lead to additional effects such as turbulent pumping
that might oppose the instability.

Extending the DNS of BKR to the case with stratification is
therefore one of the main goals of the present paper.
This will allow us to make a meaningful comparison
between DNS in a stratified fluid with mean-field modeling.
We are now also able to present data for cases in which small-scale
dynamo action is possible.
This requires that the magnetic Reynolds number is large enough.
As alluded to above, it is then advantageous to choose a scale
separation ratio that is not too extreme.
While structure formation by the negative effective
magnetic pressure phenomenon becomes impossible for small scale separation
ratios, there is then also the advantage that the analysis becomes
more straightforward in that horizontal and time averages can be employed.
This would become problematic in the presence of structures that would break
the assumptions of stationarity and homogeneity in the horizontal direction.

There are two other possible caveats that may result from the simplification
of using an isothermal equation of state with isothermal stratification.
First, the effects of convection and a convectively unstable
stratification on the mean Lorentz force are ignored.
Fortunately, those turn out to be weak, as shown in a separate paper by
\cite{Kapy11}.
Second, owing to the spatial uniformity of the forcing
function, $\overline{\uu^2}$ is nearly uniform, so the effects of
turbulent pumping down the gradient of $\overline{\uu^2}$ are ignored.
We refer to these as models of type U.
Furthermore, because of strong stratification of $\rho$, the equipartition
field strength $\Beq=(\mu_0\overline{\rho\uu^2})^{1/2}$ also varies.
This has the advantage that a single simulation with imposed field $B_0$
spans a large range in the relevant control parameter $B_0/\Beq$.
However, in view of applications to turbulence in stellar convection
this is unrealistic, because there $\overline{\rho\uu^3}$ is nearly
independent of height, so $\Beq$ increases with depth only like $\rho^{1/6}$.
For this reason we also study models in which $\Beq$ is nearly constant.
We refer to these as models of type B.
For this purpose we determine first the relevant turbulent pumping
velocity which is then used in a suitably adapted mean-field model.

We begin by discussing first the determination of turbulent
transport coefficients in \Sec{DNSmodel}, present our results in
\Sec{Results}, focusing especially on models of type U, turn then in
\Sec{ComparisonUB} to models of type B, compare the results at the
level of mean-field models, and finish with a discussion of the main
differences between the magnetic buoyancy instability and the
negative effective magnetic pressure instability (NEMPI), before
concluding in \Sec{Concl}.

\section{DNS model and analysis}
\label{DNSmodel}

We consider a cubic computational domain of size $L^3$.
The smallest wavenumber is then $k_1=2\pi/L$.
We adopt an isothermal equation of state with constant
sound speed $\cs$, so the gas pressure is $p=\rho\cs^2$.
The isothermal equation of state applies to both the
background flow (the hydrostatic equilibrium) and the fluctuating flow.
In the presence of gravity, $\grav=(0,0,-g)$, where $g$
is the constant gravitational acceleration, this leads to an exponentially
stratified density,
\EQ
\rho=\rho_0\exp(-z/H_\rho),
\EN
with a constant density scale height $H_\rho=\cs^2/g$ and a normalization
factor $\rho_0$.
For all our calculations we choose $k_1 H_\rho=1$.
This implies that the number of scale heights is $\Delta\ln\rho=L/H_\rho=2\pi$,
corresponding to a density contrast of $\exp2\pi\approx535$.
This state is also chosen as our initial condition.
Note that this is an equilibrium solution that is not affected by the
possible addition of a {\it uniform} magnetic field $\BB_0$.

We solve the equations of compressible magneto-hydrodynamics in the form
\begin{equation}
\rho{\DD\UU\over\DD t}=\JJ\times\BB-\cs^2\nab\rho+\nab\cdot(2\nu\rho\SSSS)
+\rho(\ff+\grav),
\end{equation}
\begin{equation}
{\partial\AAA\over\partial t}=\UU\times\BB+\eta\nabla^2\AAA,
\end{equation}
\begin{equation}
{\partial\rho\over\partial t}=-\nab\cdot\rho\UU,
\end{equation}
where $\nu$ and $\eta$ are respectively kinematic viscosity and magnetic diffusivity,
$\BB=\BB_0+\nab\times\AAA$ is the magnetic field consisting
of an imposed uniform mean field, $\BB_0=(0,B_0,0)$, and a nonuniform part
that is represented in terms of the magnetic vector potential $\AAA$,
$\JJ=\nab\times\BB/\mu_0$ is the current density, and
${\sf S}_{ij}=\half(\partial_i U_j+\partial_j U_i)-\onethird\delta_{ij}\nab\cdot\UU$
is the traceless rate of strain tensor.
The turbulence is driven with a forcing function $\ff$ that consists of
non-polarized random plane waves with an average wavenumber $\kf=5\,k_1$.
The forcing strength is arranged such that the turbulent rms velocity,
$\urms=\bra{\uu^2}^{1/2}$, is around $0.1\,\cs$.
This value is small enough so that compressibility effects are confined
to those associated with stratification alone.

Our simulations are characterized by several non-dimensional parameters.
We define the Reynolds number as $\Rey=\urms/\nu\kf$, the
magnetic Prandtl number as $\Pm=\nu/\eta$
and the magnetic Reynolds number as $\Rm=\Rey \, \Pm$.
We anticipate that it is important to have $\Pm<1$.
However, in order to reach somewhat larger values of $\Rm$
we now choose as our primary model $\Pm=0.5$ instead of 0.25,
as was the case in BKR.
In some additional cases, we span the entire range from $\Pm=1/8$ to $\Pm=8$.
For large enough values of $\Rm$ and $\Pm$, there is small-scale dynamo action.
We define the equipartition field strength both as a function of $z$
and for the middle of the domain, i.e.\
\EQ
\Beq(z)=(\mu_0\overline{\rho\uu^2})^{1/2},\quad
\Beqz=(\mu_0\rho_0)^{1/2}\,\urms.
\label{Beqz}
\EN
The latter will be used to specify the normalized strength of the imposed
horizontal field, which is also independent of height.
Another alternative is to normalize by the equipartition field strength
at the top of the domain.
In our models with nearly height-independent turbulent velocity,
this would make the imposed field strength normalized by the equipartition
value at the top $\approx5$ times bigger.

In all cases we adopt stress-free perfect conductor boundary conditions
at top and bottom of the domain.
The simulations are performed with the {\sc Pencil Code}%
\footnote{{\tt http://pencil-code.googlecode.com}},
which uses sixth-order explicit finite differences in space and a
third-order accurate time stepping method \citep{BD02}.

In this paper we present two groups of runs.
In the first group we have the same forcing amplitude at all
heights while in the second group we adjust the forcing such that
the rms velocity depends on height such that the turbulent
kinetic energy density is nearly independent of height.
In contrast to earlier work where it was possible to analyze the results
in terms of volume averages, we now have to restrict ourselves to
horizontal averages which show a strong dependence on height.
Thus, we determine the contribution to the mean momentum density that comes
from the fluctuating field:
\EQ
\overline\Pi_{ij}^{f}=\meanrho\,\overline{u_iu_j}
+\half\delta_{ij}\overline{\bb^2}-\overline{b_ib_j},
\label{stress0}
\EN
where the $\mu_0$ factor is dropped from now on and
overbars indicate $xy$ averages.
The superscript f signifies the contributions from the fluctuating field.
This, together with the contribution from the mean field, namely
\EQ
\overline\Pi_{ij}^{m}=\meanrho \, \meanU_i\meanU_j
+\delta_{ij}\left(\overline{p}+\half\meanBB^2\right)
-\meanB_i\meanB_j-2\nu \meanrho \, \overline{\sf S}_{ij},
\EN
comprises the total mean momentum tensor,
and the averaged momentum equation is given by:
\EQ
{\partial\over\partial t} \meanrho \, \meanUU_i =
-\nabla_j \left(\overline\Pi_{ij}^{f}+\overline\Pi_{ij}^{m}\right)
+ \meanrho \, g_i .
\EN
Here $\meanUU$ and $\meanBB$ are the mean velocity and magnetic
fields, $\overline{p}$ is the mean fluid pressure.
We are interested in the contribution to \Eq{stress0}
that arises only from the presence of the mean field, so we subtract
the corresponding tensor components that are obtained in the absence
of the mean field.
We thus define
\EQ
\Delta\overline\Pi_{ij}^{f}\equiv
\overline\Pi_{ij}^{{f},\overline{B}}-\overline\Pi_{ij}^{f,0},
\label{DeltaPi}
\EN
for which we make the following ansatz (RK07):
\EQ
\Delta\overline\Pi_{ij}^{f}=
-\left(\half\qp\delta_{ij}+\qg\ghat_i \ghat_j \right)\meanBB^2
+\qs\meanB_i\meanB_j,
\label{PifB0}
\EN
where $\gghat$ is the unit vector in the direction of gravity
and the coefficients $\qp$, $\qs$ and $\qg$ are expected to be
functions of the modulus of the field, $\meanB\equiv|\meanBB|$.
Equation (\ref{PifB0}) can also be obtained from symmetry arguments,
i.e., in the case of a horizontal imposed field, the linear
combination of three independent true tensors, $\delta_{ij}, \ghat_i \ghat_j$
and $\overline{B}_i \overline{B}_j$, yields ansatz (\ref{PifB0}).

The meaning of the turbulence coefficients $\qp$, $\qs$ and $\qg$
is as follows.
The coefficient $\qp$ represents the isotropic
turbulence contribution to the mean magnetic pressure, while
$\qg$ is the anisotropic turbulence contribution to the mean
magnetic pressure, and the coefficient $\qs$ is
the turbulence contribution to the mean magnetic tension.
In the theory of RK07, the coefficients $\qp$, $\qs$ and $\qg$ have been
obtained using the spectral $\tau$ approach and the
renormalization approach. The $\tau$ approach has been justified
in a number of numerical simulations \citep{BKM04,BS05b,BS07}.
However, if there is insufficient scale separation, higher order
terms such as $\meanJ_i\meanJ_j$ would need to be included.
For helical flows, terms involving $\meanJ_i\meanB_j$ and
$\meanB_i\meanJ_j$, could also be present.
Such terms are not included with the uniform fields used in
the present study.

The effective mean Lorentz force that takes into account the turbulence
effects, reads:
\EQ
\meanrho \, \meanFFFF^{M}_i = -\nabla_j
\Big(\half\meanBB^2 \delta_{ij} -\meanB_i\meanB_j
+ \overline\Pi_{ij}^{{f},\overline{B}}-\overline\Pi_{ij}^{f,0}\Big).
\label{Lor-force}
\EN
Except for the contribution proportional to $\ghat_i \ghat_j$
and the fact that we use here horizontal averages,
\Eq{PifB0} is equivalent to that used in BKR,
where full volume averages were used.
Asymptotic expressions for the $\meanB$ dependence of $\qp$, $\qs$,
and $\qe$ are given in \App{NonlinearCoefficients}.
Here we use DNS of density-stratified turbulence
to determine these coefficients.
In the present case, we have $\meanBB\approx(0,\meanB,0)$,
so \Eq{PifB0} yields
\EQA
\Delta\overline\Pi_{xx}^{f}
&=&-\half\qp\meanBB^2,
\nonumber\\
\Delta\overline\Pi_{yy}^{f}
&=&-(\half\qp-\qs)\meanBB^2,
\label{qp} \\
\Delta\overline\Pi_{zz}^{f}
&=&-(\half\qp+\qe)\meanBB^2,
\nonumber
\ENA
where we have computed $\Delta\overline\Pi_{ii}^{f}$ from DNS as
\EQ
\Delta\overline\Pi_{ii}^{f}
=\meanrho\,(\overline{u_i^2}-\overline{u_{0i}^2})
+\half(\overline{\bb^2}-\overline{\bb_0^2})
-(\overline{b_i^2}-\overline{b_{0i}^2}),
\EN
where no summation over the index $i$ is assumed.
The subscripts 0 indicate values obtained from a reference run with $B_0=0$.
This expression takes into account small-scale dynamo action
which can produce finite background magnetic fluctuations $\bb_0$.
(Thus, the reference run is {\em not} non-magnetic.)
The critical magnetic Reynolds number for small-scale dynamo action is
between 30 and 160, depending on the value of the magnetic Prandtl number \citep{Iska,B11}.
\EEq{qp} is then used to obtain explicit expressions for
\EQA
\qp&=&-2\Delta\overline\Pi_{xx}^{f}/\meanBB^2,
\nonumber \\
\qs&=&
(\Delta\overline\Pi_{yy}^{f}-\Delta\overline\Pi_{xx}^{f})/\meanBB^2,
\nonumber \\
\qg&=&
-(\Delta\overline\Pi_{zz}^{f}-\Delta\overline\Pi_{xx}^{f})/\meanBB^2,
\nonumber \\
\ENA
allowing $\qp$, $\qs$, and $\qg$ to be evaluated at each height $z$.

\section{Results}
\label{Results}

\subsection{Effective mean magnetic pressure}
\label{EffectiveMeanPressure}

\begin{figure}\begin{center}
\includegraphics[width=\columnwidth]{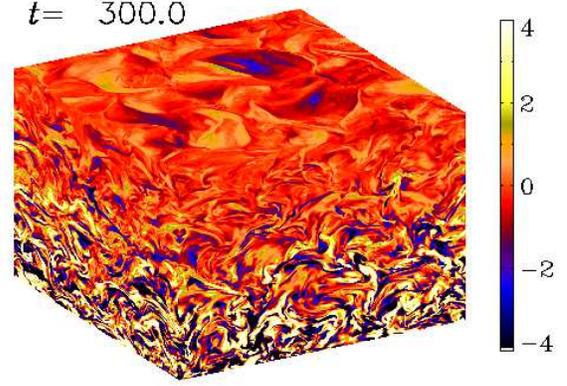}
\end{center}\caption[]{
Visualization of $\Delta B_y/\Beq$ on the periphery of the computational domain
for Model U1h600 with $B_0/\Beqz=0.1$ at $\Rm\approx600$ using $512^3$ meshpoints.
}\label{B01_kf5_512z_pm05_pc_t300}\end{figure}

We begin by considering the turbulence effects on the effective mean magnetic
pressure using a sequence of models of type U in which the rms velocity
of the turbulence intensity is approximately independent of height, so
$\Beq$ varies like $\rho^{1/2}$ and is about 23 times smaller at the top than at the bottom.
In \Fig{B01_kf5_512z_pm05_pc_t300} we show a visualization of the departure
of $B_y$ from the imposed field, $\Delta B_y\equiv B_y-B_0$,
on the periphery of the computational domain for our model with the
largest resolution (Model U1h600; for a complete list of all models
discussed in this paper see \Tab{ModelsU}).
It turns out that most of the variability of the magnetic field
occurs near the bottom of the computational domain.
This is caused by the local variation of $\Beq\propto\rho^{1/2}$.
Therefore, $B_0/\Beq$ is large in the upper parts, making it
less easy for the turbulence to produce strong fluctuations
due to the enhanced work done against the Lorentz force.
By contrast, in the lower parts, $B_0/\Beq$ is small,
allowing magnetic fluctuations to be produced.

In the following we frequently use the symbol $\beta$ to denote
normalization by $\Beq$, e.g., $\beta=\meanB/\Beq(z)$.
However, when we give the strength of the ($z$-independent) imposed or rms fields,
we normalize with respect to $\Beqz$, i.e., $\beta_0=B_0/\Beqz$
and $\betarms=B_{\rm rms}/\Beqz$.
The symbol $\beta$ used here is not to be confused with the
``plasma beta'', which denotes the ratio of gas to magnetic pressures.
To avoid confusion, we always spell out ``plasma beta'' in words.

\begin{table}[b!]\caption{
Summary of all DNS models discussed in this paper.
}\vspace{12pt}\centerline{\begin{tabular}{lccccccc}
Model & $\Rm$ & $\Pm$ & $\beta_0$ & $\qpz$ & $\betastar$ & $\betarms$ & Res.\\
\hline
U1h05 & 0.5&0.5& 0.1 & 0.2 & 0.32 &*0.04&$64^2\times128$ \\
U1h1  & 1.5&0.5& 0.1 & 0.2 & 0.32 &*0.08&$64^2\times128$ \\
U1h5  &  5 &0.5& 0.1 &  4  & 0.34 &*0.15&$64^2\times128$ \\
U1h10 & 10 &0.5& 0.1 &  13 & 0.33 &*0.20&$128^3$ \\
U1h20 & 23 &0.5& 0.1 &  40 & 0.33 &*0.29&$128^3$ \\
U1o35 & 35 & 1 & 0.1 &  90 & 0.35 &*0.40&$128^3$ \\
U1t35 & 35 & 2 & 0.1 &  70 & 0.34 &*0.44&$128^3$ \\
U1f35 & 35 & 4 & 0.1 &  15 & 0.28 &*0.44&$128^3$ \\
U1e35 & 35 & 8 & 0.1 & 0.2 & 0.32 &*0.38&$128^3$ \\
U1h40 & 42 &0.5& 0.1 & 170 & 0.38 &*0.82&$128^3$ \\
U1q70 & 70 &1/4& 0.1 & 250 & 0.37 &0.14 &$128^3$ \\
{\bf U1h70}&{\bf70}&{\bf0.5}&{\bf0.1}&{\bf100}&{\bf0.33}&{\bf0.31}&${\bf128^3}$
                                                 \\
U1h70h& 70 &0.5& 0.1 &  60 & 0.30 &0.29 &$256^3$ \\
U1o70 & 70 & 1 & 0.1 &  50 & 0.29 &0.42 &$128^3$ \\
U1t70 & 70 & 2 & 0.1 &  50 & 0.26 &0.49 &$128^3$ \\
U1f70 & 70 & 4 & 0.1 &  20 & 0.22 &0.55 &$128^3$ \\
U1e70 & 70 & 8 & 0.1 &  -- & --   &0.49 &$128^3$ \\
U2h70 & 70 &0.5& 0.2 & 130 & 0.35 &0.31 &$128^3$ \\
U5h70 & 70 &0.5& 0.5 & 200 & 0.39 &0.31 &$128^3$ \\
U1a140&140 &1/8& 0.1 & 200 & 0.31 &0.43 &$256^3$ \\
U1q140&140 &1/4& 0.1 &  40 & 0.27 &0.49 &$128^3$ \\
U1h140&140 &0.5& 0.1 &  50 & 0.27 &0.53 &$128^3$ \\
U1h250&250 &0.5& 0.1 &  40 & 0.20 &0.68 & $256^3$ \\
U1h600&600 &0.5& 0.1 &  40 & 0.22 &0.82 & $512^3$ \\
B07h35& 35 &0.5& 0.07&  -- &  --  &*0.15& $128^3$\\
B2h35 & 35 &0.5& 0.2 &  -- &  --  &*0.30& $128^3$ \\
\label{ModelsU}\end{tabular}}
\footnotetext{Here, $\beta_0=B_0/\Beqz$ and $\betarms=B_{\rm rms}/\Beqz$
denote field strengths in equipartition units, while $\betastar$
is a fit parameter that applies locally.
Normally, $\betarms=B_{\rm rms}/\Beq$ refers to the field generated
by small-scale dynamo action in the reference run with $B_0=0$,
except when there is an asterisk indicating that there is {\it no} small-scale dynamo
and $\betarms$ gives the result from tangling of the applied field
in the corresponding run with $B_0\neq0$.
Our reference model is indicated in bold face.
}\end{table}

\begin{figure}\begin{center}
\includegraphics[width=\columnwidth]{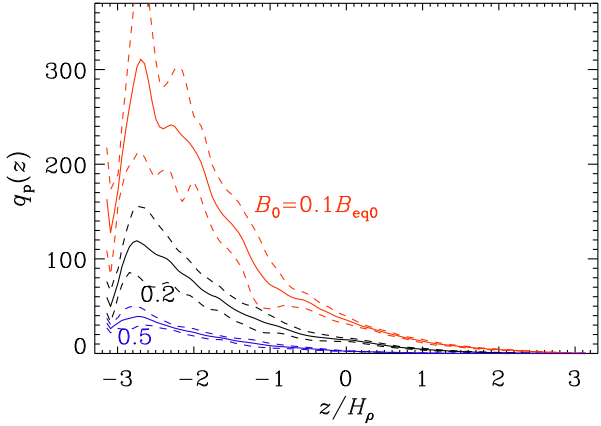}
\end{center}\caption[]{
Dependence of $q_p(z)$ (solid lines) with error margins (dashed lines)
as functions of $z$, for Models U1h70, U2h70, and U5h70 with
$B_0=0.1\Beqz$, $B_0=0.2\Beqz$, and $B_0=0.5\Beqz$, with $\Rm\approx70$,
$g/\cs^2k_1=1$, and a density contrast of 530.
Note that $q_p(z)$ reaches a maximum at the bottom of the domain
where $B_0/\Beq(z)$ is minimal.
}\label{pxyaver_compb}\end{figure}

We have computed $\qp$ for all models of type U.
We plot in \Fig{pxyaver_compb} the dependence of $\qp$ on height
for three different values of $B_0$.
In the following, the case with $B_0/\Beqz=0.1$ will be used
as our fiducial run.
To improve the statistics, we present here time averaged results
of $\qp$, which itself is already averaged over $x$ and $y$.
Error bars have been calculated by dividing the time series into
three equally long pieces and computing the maximum departure
from the total average.
In agreement with earlier work, $\qp$ is always positive and
exceeds unity when the mean magnetic field is not sufficiently strong.
This is the case primarily at the bottom of the domain
(negative values of $z$) where the density is high
and therefore the magnetic field, in units of
the equipartition field strength, is weak.
Since $B_0=\const$ and $\Beq$ increases with depth,
$B_0/\Beq$ is smallest at the bottom, so $\qp$ also increases.
The sharp uprise toward the lower boundary is just a result of the exponential
increase of the density combined with the fact that the horizontal
velocity reaches a local maximum on the boundary.

\begin{figure}\begin{center}
\includegraphics[width=\columnwidth]{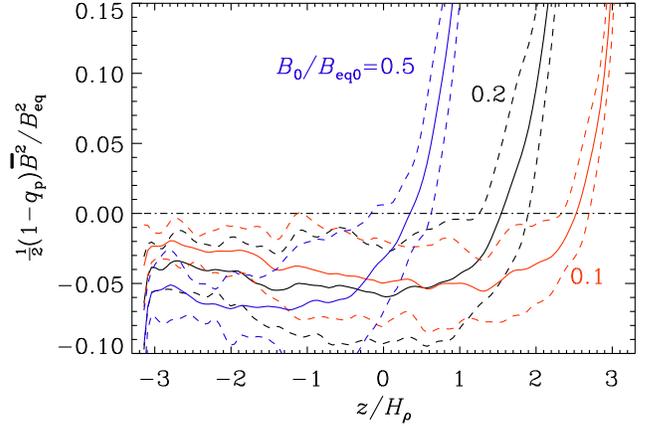}
\end{center}\caption[]{
Normalized effective mean magnetic pressure as a function of depth
for the same models as in \Fig{pxyaver_compb}.
Note that this function now reaches a negative minimum somewhere in
the middle of the domain.
}\label{pxyaver_compp}\end{figure}

The total effective magnetic pressure of the mean field
(that takes into account the effects of turbulence
on the mean Lorentz force) is given by
$\half[1-\qp(\meanB)]\meanBB^2$.
This has to be compared with the turbulent
kinetic energy density,
$\half\overline{\rho\uu^2}$.
Small contributions of terms $ \propto \qe$ to the effective
mean magnetic pressure are discussed in Sect. 3.3.
In \Fig{pxyaver_compp} we plot the effective magnetic pressure
normalized by $\Beq^2$,
\EQ
\Peff=\half(1-\qp)\meanBB^2/\Beq^2,
\EN
where $\Beq^2$ itself is a function of height; see \Eq{Beqz}.
It turns out that this function now reaches a negative minimum somewhere
in the middle of the domain.
Work of \cite{KBKR} has shown that the regions below the minimum value
of $\Peff$ are those that can potentially display NEMPI.

\begin{figure}\begin{center}
\includegraphics[width=\columnwidth]{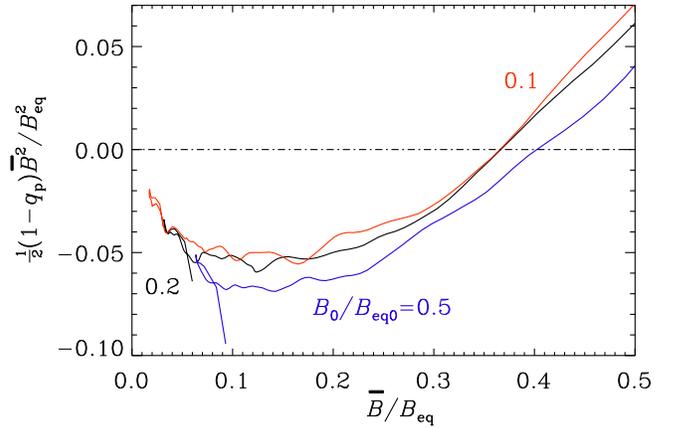}
\end{center}\caption[]{
Same as \Fig{pxyaver_compp}, but as a parametric representation
as function of the local value of the ratio $B_0/\Beq(z)$.
Note that the curves for $B_0=0.1\Beqz$, $B_0=0.2\Beqz$, and $B_0=0.5\Beqz$
collapse onto a single dependency.
The error range is the same as in the previous figure,
but not shown for clarity.
}\label{pxyaver_compp_vs_B}\end{figure}

\begin{figure}\begin{center}
\includegraphics[width=\columnwidth]{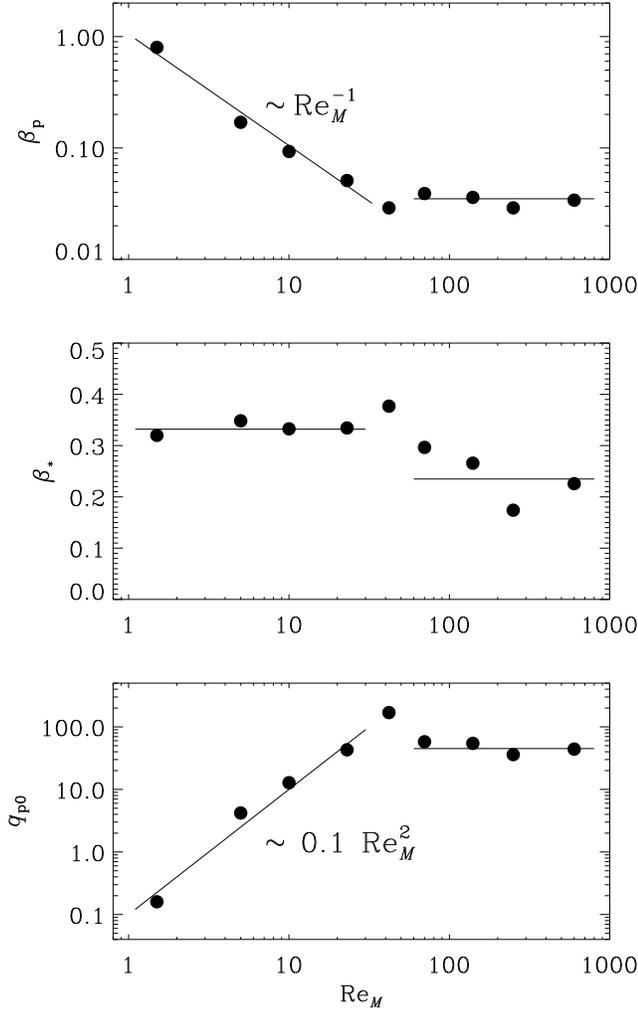}
\end{center}\caption[]{
Dependence of the fit parameters on $\Rm$ using Models U1h05--U1h600.
}\label{pfitR}\end{figure}

We expect that $\qp$ is a function of the ratio of $\meanB/\Beq$.
This was observed numerically in BKR for constant $\Beq$ by varying
the value of $B_0$ to obtain $\qp$ for a range of different simulations.
In the present case, however, $\Beq$ is a function of $z$, which
is the main reason why $\qp$ depends on height.
In \Fig{pxyaver_compp_vs_B} we plot the effective mean magnetic pressure
as a function of magnetic field in units of the local equipartition value.
Note that now all three curves for different values of $\meanB$ collapse
onto a single curve, which demonstrates that the dependence of $\qp$
on both $\meanB$ and $z$ can indeed be reduced to a single dependence
on the ratio $\beta=|\meanBB|/\Beq(z)$.

To quantify the form of the $\qp(\beta)$ dependence, we used in BKR
a fit formula involving an arctan function.
However, following recent work of \cite{KBKR}, a sufficient and
certainly much simpler fit formula is
\EQ
\qp(\beta)={\qpz\over1+\beta^2/\betap^2}.
\label{qpfit}
\EN
Here the fit parameters $\qpz$ and $\betap$ are determined
by measuring the minimum effective magnetic pressure, $\Pmin=\min(\Peff)$,
as well as the position of the minimum, $\Bmin$, where $\Peff(\Bmin)=\Pmin$.
For our setups, $\Pmin$ is typically around $-0.05$, while
$\betamin=\Bmin/\Beq$ is between 0.1 and 0.2.
This is remarkably close to Fig.~3 of RK07, who used
the spectral $\tau$ relaxation approximation.
The fact that nearly the same functional form for the effective magnetic
pressure of the mean field is obtained, supports the idea that this
effect is robust.

For many of the models in \Tab{ModelsU} we have determined the fit parameters
$\qpz$ and $\betap$.
It turns out that for small values of $\Rm$, $\qpz$ increases quadratically
with $\Rm$ and $\betap$ decrease like $\Rm^{-1}$.
Thus, for $\Rm<30$, we have $\betastar^2\equiv\betap^2\qpz\approx\const$.
The significance of this is that $\betastar$ turns
out to be nearly independent
of $\Rm$; see \Fig{pfitR}.
It allows rewriting the fit formula as
\EQ
\qp(\beta)={\beta_*^2\over\betap^2+\beta^2},
\EN
where for small values of $\Rm$, only $\betap$ depends on $\Rm$.
In particular, we have then $\betap\approx1.05/\Rm$.

The obtained scaling for $\qpz \sim 0.1 \Rm^2$ is consistent
with an estimate based on the quasi-linear calculations.
This analysis is similar to that of \cite{RKS12}, except that we
performed an explicit integration in $\omega {\bm k}$-space for a
power-law kinetic energy spectrum of the background turbulence
and for a Lorentz profile for the frequency dependence of
the velocity correlation function.
For $\beta \ll 1$ this analysis yields the expression (see Appendix~\ref{SOCA})
\EQ
\qpz={8-\Pm\over60\,\Pm}\Rm^2,
\label{qpz}
\EN
where $\Rm^2/\Pm=\Rm\Rey$.
For $\Pm=1/2$, we have $\qp=0.25\,\Rm^2$, which is in qualitative
agreement with our scaling for $\qpz$.
The discrepancy in the coefficient is related to the fact that
the quasi-linear approach is only
valid for small magnetic and fluid Reynolds numbers.
Therefore, the limit $\Pm \to 0$ in the framework of
the quasi-linear approach only implies the case of large
magnetic diffusion $\eta$, while the case of small
$\nu$ need to be considered in the framework of
approaches that are valid for large fluid Reynolds numbers
(like the $\tau$ relaxation approach).

Looking at \Tab{ModelsU}, it may seem surprising that $\qpz$ can reach values
as large as 250 (see, e.g., Model U1q70 with $\Rm\approx250$ and $\Pm=4$).
However, the more relevant quantity is $\qp(\beta)\beta^2$, which is of
the order of $\beta_*^2$ for large field strengths.
Summarizing the results from  \Fig{pfitR}, we find
\EQ
\betap\approx1.05\,\Rm^{-1},\quad\betastar=0.33\quad\mbox{(for $\Rm<30$)},
\label{loRm}
\EN
and
\EQ
\betap\approx0.035,\quad\betastar=0.23\quad\mbox{(for $\Rm>60$)},
\label{hiRm}
\EN
which implies that $\qp(\beta)\beta^2$ is below 0.1 and 0.06 for the regimes
applicable to \Eqs{loRm}{hiRm}, respectively.
Also, it is tempting to associate the sudden drop of $\betastar$
from 0.33 to 0.23 with the onset of small-scale dynamo action.
The fact that small-scale dynamo action reduces the negative effect of
turbulence on the effective mean magnetic pressure was already predicted
by RK07 and is also quite evident by looking at \Tab{ModelsU}, where
$\qpz$ is found to reach more moderate values after having reached
a peak at $\Rm\approx30$.

We reiterate that, as long as the value of the plasma beta
(i.e.\ the ratio of gas pressure to magnetic pressure),
is much larger than unity, our results are independent of
the plasma beta.
What matters is the ratio of magnetic energy
density to kinetic energy
density, not the thermal energy density.
This is also clear from the equations given in \App{NonlinearCoefficients}.
In the present simulations, the plasma beta varies from between 5 and
100 at the top to around  $10^5$ at the bottom, so the total pressure
(gas plus magnetic plus turbulent pressure) is always positive.

\subsection{Dependence on magnetic Prandtl number}

\begin{figure}\begin{center}
\includegraphics[width=\columnwidth]{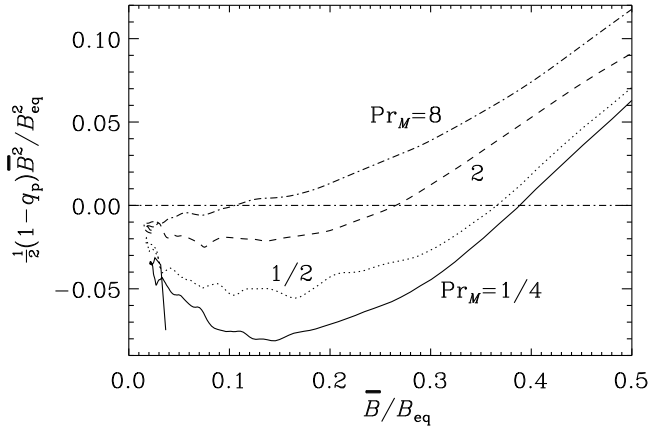}
\end{center}\caption[]{
Normalized effective mean magnetic pressure for different values of $\Pm$,
for Models U1q70, U1h70, U1t70, and U1e70,
where $\Rm\approx70$ and $B_0=0.1\Beqz$.
}\label{prandtlsweep}\end{figure}

In most of the runs discussed above we used $\Pm=1/2$.
As expected from earlier work of RK07, the negative magnetic pressure
effect should be most pronounced at small $\Pm$.
This is indeed confirmed by comparing with larger and smaller values
of $\Pm$; see \Fig{prandtlsweep}, where we show the results
for Models U1q70, U1h70, U1t70, and U1e70.
Here and in \Tab{ModelsU} the different values of $\Pm$ are denoted
by the letters q, h, o, t, f, and e, which stand for a quarter,
half, one, two, four, and eight, while the letter a is used for one eighth.

\subsection{Resolution dependence}

\begin{figure}\begin{center}
\includegraphics[width=\columnwidth]{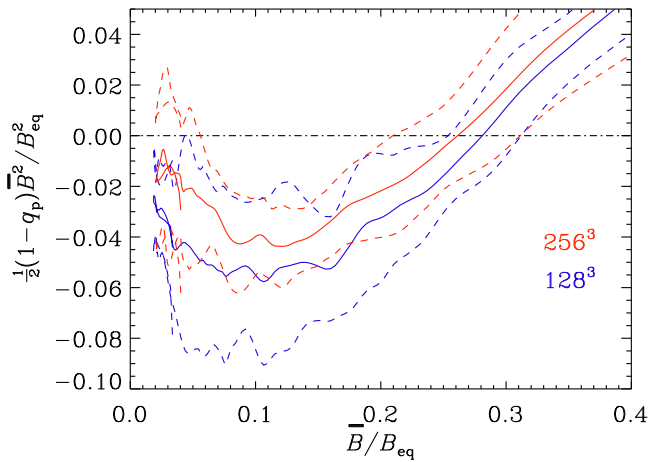}
\end{center}\caption[]{
Resolution dependence of $\qp(\meanB_y/\Beq)$ for $\Pm=1/2$, $\Rm\approx70$
for Models U1h70 and U1h70h using $128^3$ and $256^3$ mesh points.
Error bars are marked by the dashed lines.
}\label{resolution}\end{figure}

A density contrast of over 500 may seem rather large.
However, this impression may derive
from experience with polytropic models \citep[see, e.g.,][]{Cat91}, where most of the density
variation occurs near the surface.
In our isothermal model, the scale height is constant,
so the logarithmic density change is independent of height.
\FFig{resolution} shows that the error bars for the $256^3$ run are smaller
than those for the $128^3$ run, and that the minimum of $\Peff$ is somewhat
less shallow, but within error bars the two curves are still comparable.
Here, the relevant input data are averaged over a time interval $\Delta t$
such that $\urms\kf\Delta t$ is at least 1500, and that error bars, which
are based on 1/3 of that, cover thus at least 500 turnover times.

\subsection{Coefficients $\qs$ and $\qe$}
\label{qs_and_qe}

Depending on the size and magnitude of the coefficients $\qs$ and $\qe$,
their effect on the instability could be significant.
Most importantly, a positive value of $\qs$ was found to be chiefly
responsible for producing three-dimensional mean-field structures
\citep{KBKR}, i.e., structures that break up in the direction of
the imposed field.
The coefficient $\qe$, on the other hand, affects the negative
effective magnetic pressure and could potentially enhance its
effect significantly (RK07).

\begin{figure}\begin{center}
\includegraphics[width=\columnwidth]{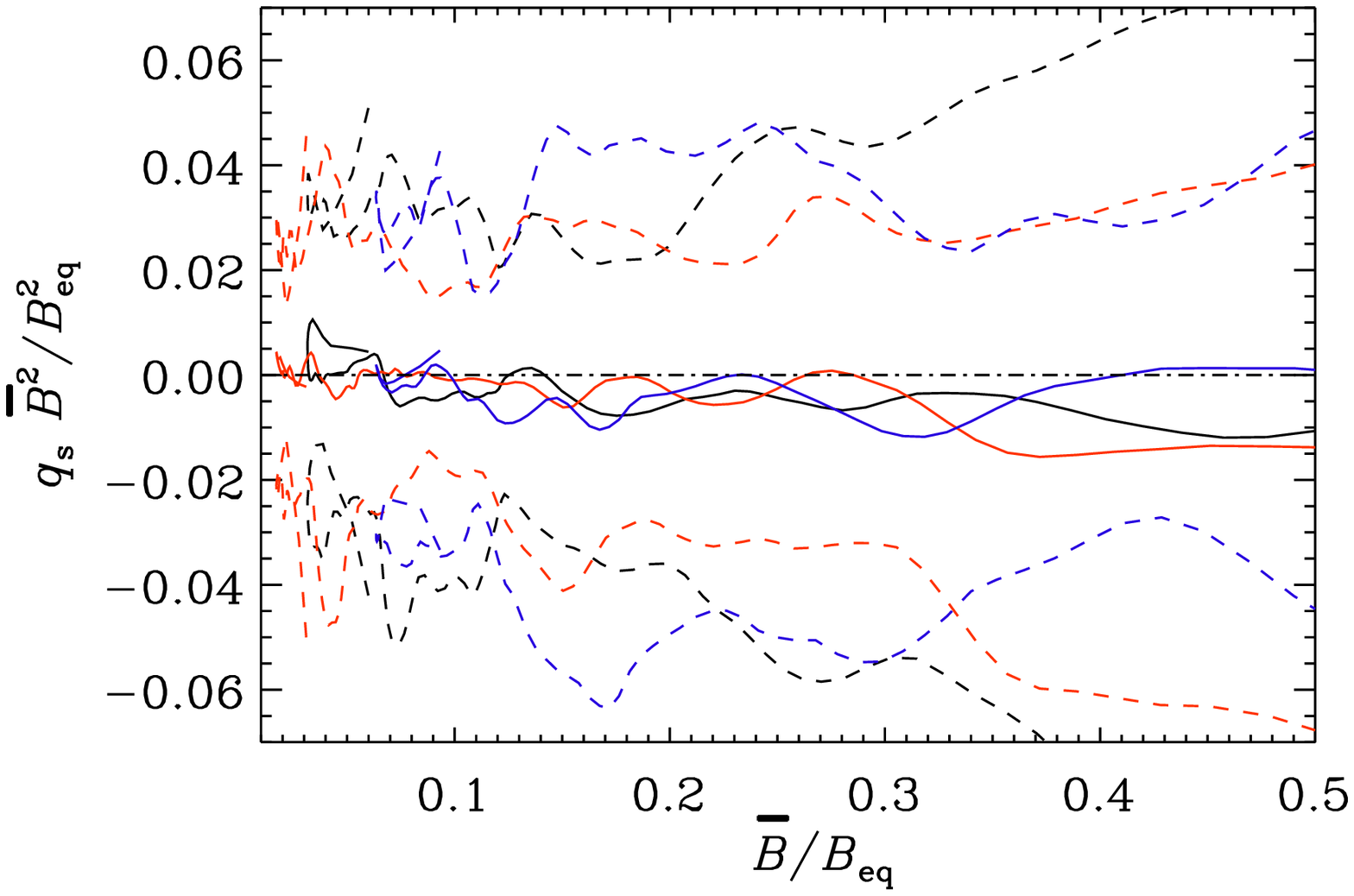}
\end{center}\caption[]{
Similar to \Fig{pxyaver_compp_vs_B},
but for $\qs\meanBB^2/\Beq^2$ and now as a function of $\meanB_y/\Beq$.
Within the error range (dashed lines), $\qs=0$ for all field strengths.
}\label{pxyaver_coms_vs_B}\end{figure}

Using Eqs.~(\ref{qp}), we now determine $\qs$ and $\qe$.
The results are shown in \Figs{pxyaver_coms_vs_B}{pxyaver_compe_vs_B}
for the three imposed field strengths considered above,
where $\overline{\uu^2}$ is nearly
independent of $z$, so $\Beq(z)$ varies by a factor of $\exp\pi\approx23$,
allowing us to scan the dependence on $\meanB/\Beq$ in a single run.
It turns out that both $\qs$ and $\qe$ are around zero.

\begin{figure}\begin{center}
\includegraphics[width=\columnwidth]{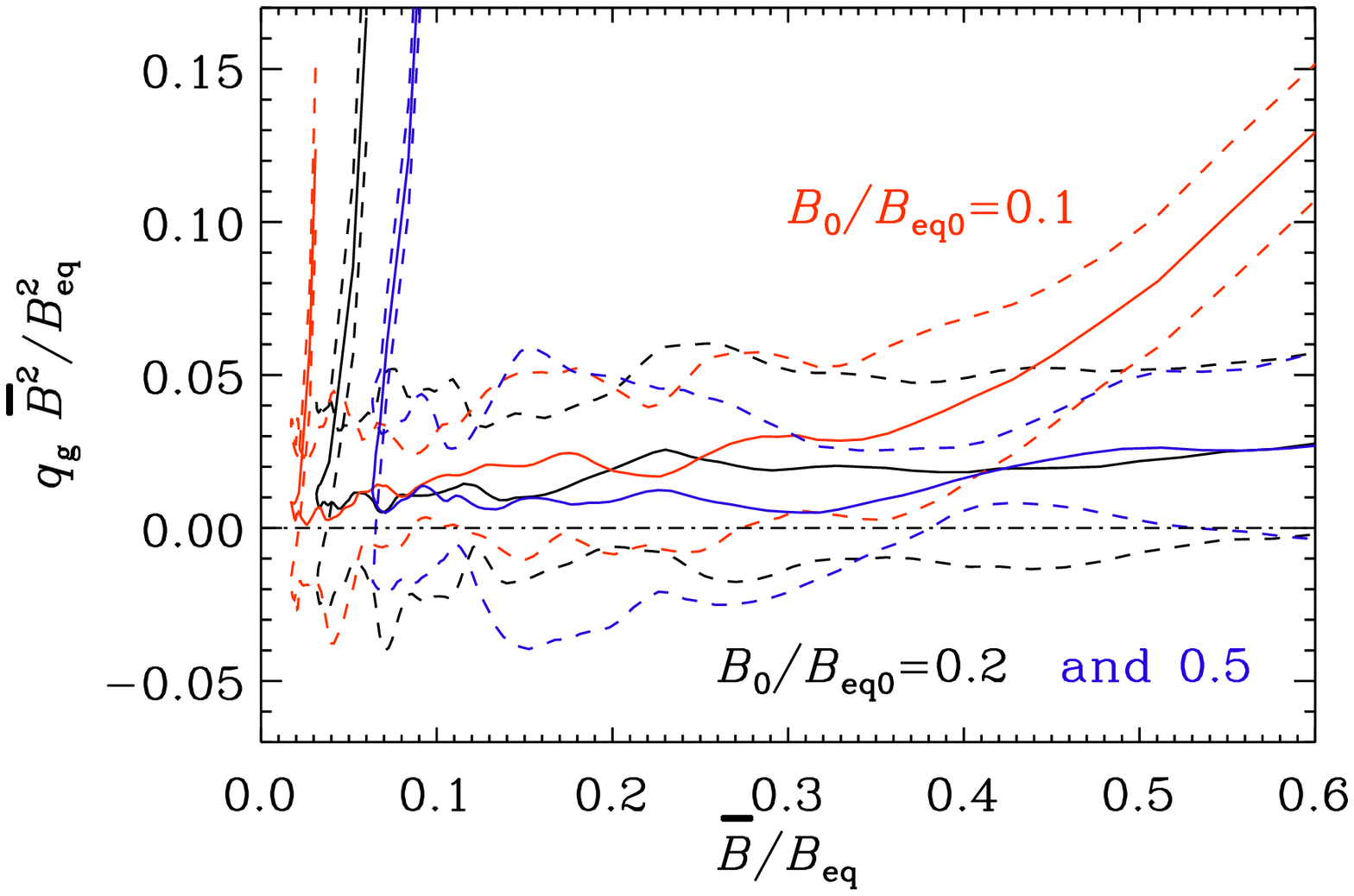}
\end{center}\caption[]{
Similar to \Fig{pxyaver_coms_vs_B}, but for $\qe$.
Note that $\qe$ is positive.
}\label{pxyaver_compe_vs_B}\end{figure}

\begin{figure*}\begin{center}
\includegraphics[width=\textwidth]{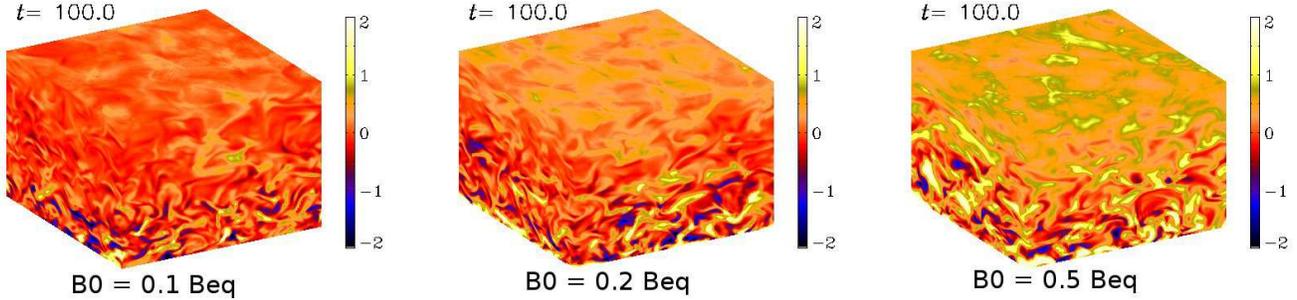}
\end{center}\caption[]{
Visualization of $\Delta B_y/\Beq$ on the periphery of the computational domain
for runs with $B_0/\Beqz=0.1$, 0.2, and 0.5, with $\Rey=70$.
}\label{Bkf5_128z_pm05_pc_499}\end{figure*}

Recent DNS of stratified convection with
an imposed horizontal magnetic field did actually yield non-vanishing
(positive) values of $\qe$ for stratified convection \citep{Kapy11}.
In the present study with vertical density stratification,
$\qe$ is much smaller, but generally positive.
This appears to be in conflict with the theoretical expectation for $\qe$
given in \App{NonlinearCoefficients}, where $\qe=O(\ellf^2/H_\rho^2)$
if we assume $\ellf=2\pi/\kf\approx1.3$, which gives $\qe\approx1.2$.
However, without the $2\pi$ factor, we would have $\ellf=\kf^{-1}=0.2$,
and thus $\qe\approx-0.03$.
\Fig{pxyaver_compe_vs_B} suggests a positive value of similar magnitude.
This issue will hopefully be clarified soon in future work.

Next, we discuss the  results for $\qs$.
In BKR there was some evidence that $\qs$ can become
positive in a narrow range of field strengths, but the error
bars were rather large.
The present results are more accurate and suggest that
$\qs\meanBB^2/\Beq^2$ is essentially zero.
This is also in agreement with recent convection simulations \citep{Kapy11}.

In summary, the present simulations provide no evidence that the
coefficients $\qs$ and $\qe$ could contribute to the large-scale
instability that causes the magnetic flux concentrations.
This is not borne out by the analytic results given in
\App{NonlinearCoefficients}.
The results from recent convection simulations fall in between the analytic
and numerical results mentioned above, because in those $\qg$ was found to be
positive, while $\qs$ was still found to be small and negative \citep{Kapy11}.

\section{Comparison of models of types U and B}
\label{ComparisonUB}

\subsection{Results from DNS}

As discussed in the beginning of \Sec{EffectiveMeanPressure},
most of the variability of the magnetic field
occurs near the bottom of the computational domain.
This is also evident from \Fig{Bkf5_128z_pm05_pc_499}, where we show
visualizations of the $y$ component of the departure from the imposed
field, $\Delta B_y$, on the periphery of the domain
for Models U1h70, U2h70, and U5h70 with $B_0/\Beqz=0.1$, 0.2, and 0.5,
respectively.

\begin{figure}\begin{center}
\includegraphics[width=\columnwidth]{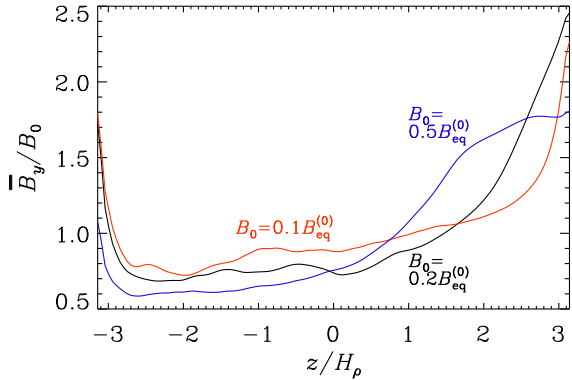}
\end{center}\caption[]{
Normalized mean magnetic field in the direction of the imposed field
versus height for $B_0=0.1\Beqz$, $0.2\Beqz$, and $0.5\Beqz$ with $\Rey=70$.
}\label{pxyaver_comp_B}\end{figure}

\begin{figure}\begin{center}
\includegraphics[width=\columnwidth]{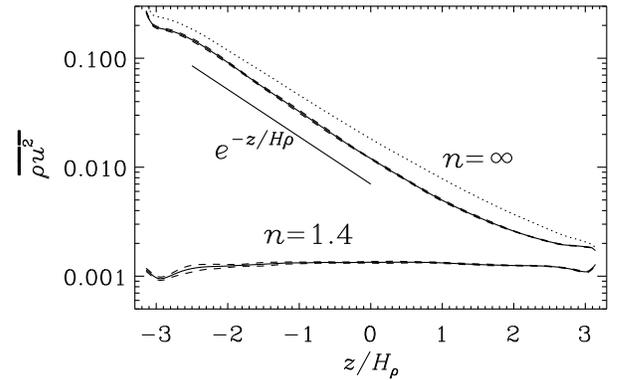}
\end{center}\caption[]{
Turbulent kinetic energy density versus height for $n=\infty$
for $B_0=0$ (dotted line) and $B_0=0.2\Beqz$
compared with the case for $n=1.4$ and $B_0=0.2\Beqz$.
}\label{pxyaver_rhou2}\end{figure}

The vertical dependence of the horizontally averaged mean magnetic field
(now normalized to $B_0=\const$) is shown in \Fig{pxyaver_comp_B}.
We see that, especially for intermediate field strengths,
there is an increase of the magnetic field near the top of the domain.
One possibility is that this is caused by the effect of nonlinear
turbulent pumping, which might cause
the mean field to be pumped
up due to the gradients of the mean turbulent kinetic energy density
in the presence of a finite mean magnetic field \citep[cf.][]{RK06}.
This type of pumping is different from the regular pumping down the
gradient of turbulent intensity \citep{Rad69}.
To eliminate this effect, we have produced
additional runs where the kinetic energy density is approximately
constant with height.
This is achieved by modulating the forcing function by a $z$-dependent
factor $e^{z/H_{\it f}}$.
We define $n=H_{\it f}/H_\rho$ and find that for $n=1.4$
the kinetic energy density is
approximately independent of height; see \Fig{pxyaver_rhou2}.

\begin{figure}\begin{center}
\includegraphics[width=\columnwidth]{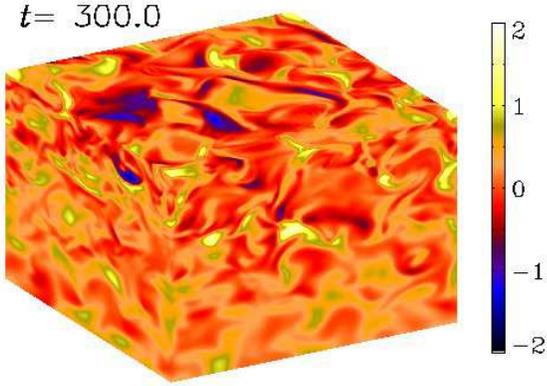}
\end{center}\caption[]{
Visualization of $\Delta B_y/\Beq$ on the periphery of the computational domain
for Model B1h35 with nearly uniform turbulent kinetic energy density using
$H_{\it f}=nH_\rho$ with $n=1.4$.
}\label{B01e_kf5_128z_pm05_pc_expzH}\end{figure}

\begin{figure}\begin{center}
\includegraphics[width=\columnwidth]{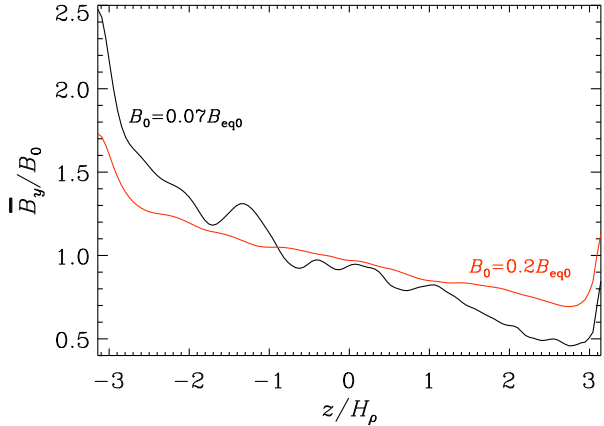}
\end{center}\caption[]{
Normalized mean magnetic field in the direction of the imposed field
versus height in the case of nearly constant turbulent kinetic energy
density, i.e.\ $\Beq(z)\approx\const$, for Models B07h35 and B2h35
with $B_0/\Beqz=0.07$ and 0.2, respectively.
}\label{pxyaver_comp_B_expzH}\end{figure}

As a consequence of reducing the turbulent driving in the lower
parts by having $\Beq(z)\approx\const$, we allow the magnetic field
to have almost the same energy density as the turbulence, i.e.\
$B_0/\Beq(z)$ is approximately independent of $z$.
This also means that the fluctuations are now no longer so pronounced
at the bottom of the domain (\Fig{B01e_kf5_128z_pm05_pc_expzH}),
where $\Rey$ drops to values around 5
and the flow is no longer turbulent.
However, at the top the Reynolds number is around 120, so here the flow
is still turbulent.
In \Fig{pxyaver_comp_B_expzH} we show the vertical dependence of the
horizontally averaged mean magnetic field in units of the imposed field
strength.
Note that now the field shows an increase toward the bottom of the domain.
This effect might be related to regular turbulent pumping \citep{Rad69},
which now has a downward component because $\overline{\uu^2}$ decreases
with depth.

\subsection{Determination of $\etat$ and $\gamma$ from the simulations}
\label{Determination}

We use the test-field method of \cite{Sch05,Sch07} in the
Cartesian implementation, as described by \cite{BRRK08},
to compute $\etat$ and $\gamma$ from the simulations
in the presence of the applied field.
We refer to this as the quasi-kinematic test-field method, which
is applicable if the magnetic fluctuations are just a consequence of
the mean field; see \cite{RB10} for details and extensions to a fully
nonlinear test-field method.
For further comments regarding the test-field method see \App{TFM}.
We analyze the two setups discussed above, namely those of
type U (where $\urms$ and hence $\etat$ are nearly constant in height)
and those of type B (where $\Beq$ is nearly constant in height).

\begin{figure}\begin{center}
\includegraphics[width=\columnwidth]{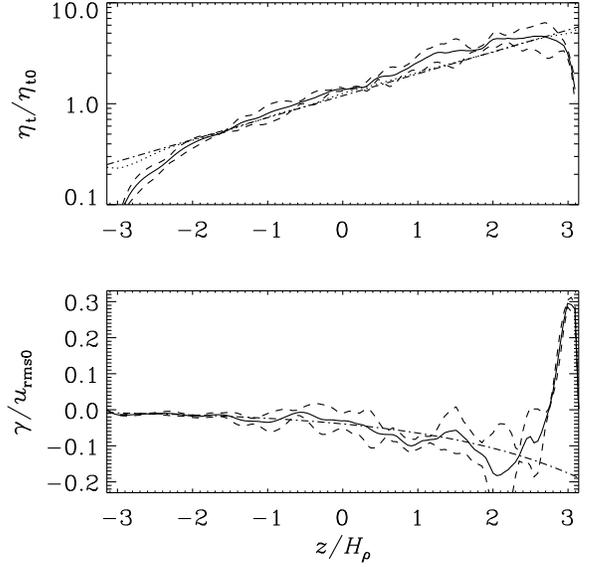}
\end{center}\caption[]{
Results for $\etat$ and $\gamma$ with the test-field method
(solid lines; error margins are shown as dashed lines) for Model B07h35.
In the upper panel, the dotted line gives $1.2\,\urms/3\kf$ and the
dash-dotted line represents $1.2\,\urmsz\exp(z/H_u)/3\kf$.
In the lower panel, the dash-dotted lines represents
$-1.2\,\urmsz\exp(z/H_u)/3\kf H_u$.
}\label{ppump}\end{figure}

The set of test fields includes constant and linearly growing ones.
For both models we use $B_0=0.01\rho^{1/2}\cs$, corresponding to
$B_0\approx0.1\Beqz$ for models of type U and $B_0\approx0.07\Beqz$ for
model of type B.
The results are shown in \Figs{ppump}{ppump_B01_kf5_128z_pm05_pc_testfield}
for Models B07h35 and U1h70, respectively.
In \Tab{Tsummary} we summarize the relevant parameters inferred for
these models.
It turns out that the DNS results are well described
by $\etat=1.2\,\etatz$, with $\etatz=\urms/3\kf$ and
$\urms(z)=\urmsz\exp(z/H_u)$, but $\gamma=-\dd\etat/\dd z$,
i.e.\ without the 1/2 factor expected from the kinematic mean-field theory
\citep{Rad69}.

\begin{figure}\begin{center}
\includegraphics[width=\columnwidth]{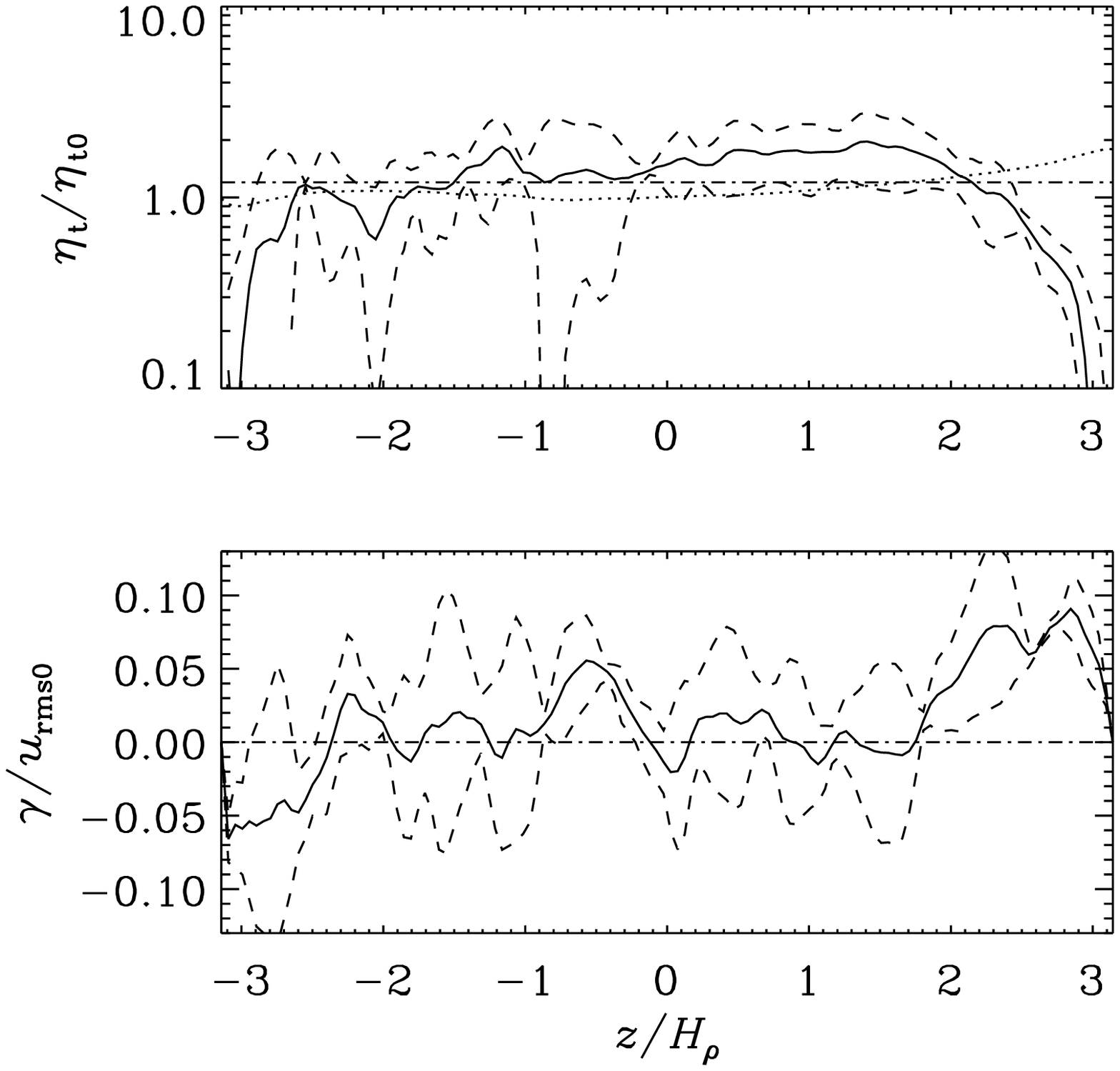}
\end{center}\caption[]{
Same as \Fig{ppump}, but for Model U1h70.
}\label{ppump_B01_kf5_128z_pm05_pc_testfield}\end{figure}

We emphasize that $\ggamma\approx\bm{0}$ for models of type U, suggesting
that additional effects owing to the mean magnetic field such as
mean-field magnetic buoyancy \citep{Kitch94,RK06}
are weak (\App{MeanFieldBuoyancy}).

\subsection{Comparison with mean-field models}
\subsubsection{Basic equations}

We follow here the same procedure as BKR and consider the equations
for the mean velocity $\meanUU$, the mean density $\meanrho$,
and the mean vector potential $\meanAA$ in the form
\EQ
{\partial\meanUU\over\partial t}=-\meanUU\cdot\nab\meanUU
-\cs^2\nab\ln\meanrho+\grav+\meanFFFF^{M}+\meanFFFF_{K,{\rm tot}},
\label{dUmeandt}
\EN
\EQ
{\partial\meanAA\over\partial t}=\meanUU\times\meanBB
+\meanEMF-\eta\meanJJ-\nab\meanPhi,
\EN
\EQ
{\partial\meanrho\over\partial t}=-\nab\cdot\meanrho\meanUU,
\label{drhomeandt}
\EN
where $\meanPhi$ is the mean electrostatic potential,
$\meanBB=\BB_0+\nab\times\meanAA$ is the mean
magnetic field including the imposed field, and
\EQ
\meanrho\meanFFFF^{M}=\meanJJ\times\meanBB+\half\nab(\qp\meanBB^2)
\EN
is the effective mean Lorentz force,
where we use for $\qp(\meanB)$ the fit formula given by \Eq{qpfit}.
However, in view of the results of \Sec{qs_and_qe},
the $\qs$ and $\qe$ terms will now be omitted, and
\EQ
\meanFFFF_{K,{\rm tot}}=(\nut+\nu)\left(\nabla^2\meanUU+\nab\nab\cdot\meanUU
+2\meanSSSS\nab\ln\meanrho\right)
\EN
is the total (turbulent and microscopic) viscous force,
\EQ
\meanEMF=\ggamma\times\meanBB-\etat\meanJJ,
\label{EMFmeanfield}
\EN
is the mean electromotive force, where $\ggamma$ is the turbulent
pumping velocity and $\etat$ is the turbulent magnetic diffusivity.
In our mean-field models we assume $\nut/\etat=1$ \citep{YBR03}.
The kinematic theory of \cite{RS75} and others predicts that
$\etat(z)=\urms(z)/3\kf$ and $\ggamma=-\half\nab\etat$.
It is fairly easy to assess the accuracy of these expressions by computing
turbulent transport coefficients from the simulations using the
test-field method \citep{Sch05,Sch07}.

A comment regarding $\meanPhi$ is here in order.
It is advantageous to isolate a diffusion operator of the
form $\etat\nabla^2\meanAA$ by using the so-called resistive gauge
in which $\meanPhi=-\etat\nab\cdot\meanAA$.
This means that the diffusion operator now becomes
$\etat\nabla^2\meanAA+(\nab\cdot\meanAA)\nab\etat$ \citep{DSB02}.
This formulation is advantageous in situations in which $\etat$ is non-uniform.

\subsubsection{Results from the mean-field models}

Next we consider solutions of \Eqss{dUmeandt}{EMFmeanfield}
for models of types U and B using the parameters specified in \Tab{Tsummary}.
To distinguish these mean-field models from the DNS results, we denote
them by script letters ${\cal U}$ and ${\cal B}$.
We have either constant $\etat$ (Models~${\cal U}$)
or constant $\Beq$ (Models~${\cal B}$).
In both cases we use $\eta=2\nu=4\times10^{-4}\cs/k_1$,
$B_0=0.005\rho^{1/2}\cs$, with $\qpz=40$ and $\betap=0.07$,
while $\Beqz$ and $\rho_0$ are given in \Tab{Tsummary} and
correspond to values used in the DNS.
This gives the profile of $\urms(z)=\Beq/\sqrt{\rho}$, which allows
us to compute $\etat(z)=\urms(z)/3\kf$ and $\nut(z)=\etat(z)$
for an assumed value of $\kf$.
In the DNS presented here we used $\kf/k_1=5$ and did not find
evidence for NEMPI, but the DNS of \cite{BKKMR11} and \cite{KBKMR}
for $\kf/k_1=15$ and $\kf/k_1=30$, respectively
did show NEMPI, so we mainly consider the case
$\kf/k_1=15$, but we also consider $\kf/k_1=5$ and 10.

As in BKR, \Eqss{dUmeandt}{drhomeandt} exhibit a linear instability with
subsequent saturation.
However, this result is still remarkable because there are
a number of differences compared with the models studied in BKR.
Firstly, we consider here an isothermal atmosphere which is stably
stratified, unlike the isentropic one used in BKR, which was only
marginally stable.
This underlines the robustness of this model and shows that this
large-scale instability can be verified over a broad range of conditions.
Secondly, this instability also works in situations where $\etat$
and/or $\Beq$ are non-uniform and where there is a pumping effect
that sometimes might have a tendency to suppress the instability.

\begin{figure}\begin{center}
\includegraphics[width=\columnwidth]{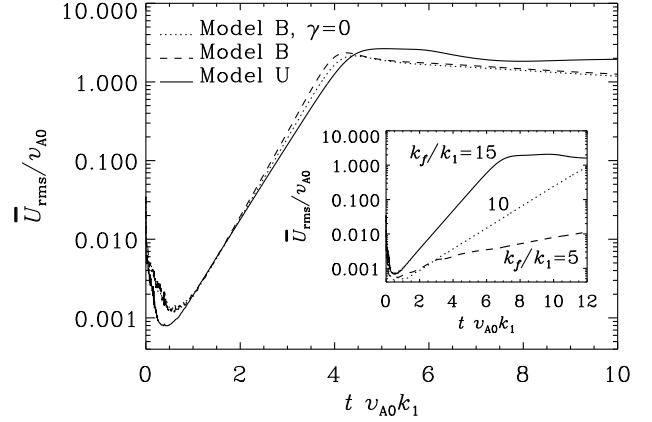}
\end{center}\caption[]{
Evolution of the mean velocity for Models~${\cal U}$ and ${\cal B}$
obtained by solving the mean field equations.
``Model~${\cal B}$0 ($\gamma=0$)'' refers to a model where
the pumping velocity is ignored.
In all cases, $\kf/k_1=15$ is assumed.
In the inset we compare the evolution for Model~${\cal U}$ with those for
smaller values of $\kf/k_1$ (Model~${\cal U}$10 and ${\cal U}$5).
}\label{pcomp}\end{figure}

In \Fig{pcomp} we compare the evolution of the rms velocity of the mean flow.
Note that, in contrast to the corresponding plots in BKR,
we have here normalized $\meanU_{\rm rms}$
with respect to $v_{A0}\equiv B_0/\sqrt{\rho_0}$
and time is normalized with respect to the Alfv\'en wave traveling
time, $(v_{A0} k_1)^{-1}$.
This was done because in these units the curves for Models~${\cal U}$
and ${\cal B}$ show similar growth rates.
This is especially true when the pumping term is ignored in Model~${\cal B}$0,
where we have set artificially $\gamma=0$.
With pumping included
(as was determined from the kinematic test-field method),
the growth rate is slightly smaller (compare dashed and dotted lines).
The pumping effect does not significantly affect the nonlinear saturation
phase, i.e.\ the late-time saturation behavior for the two versions
of Model~${\cal B}$ is similar.
Instead, the saturation phase is different for Model~${\cal U}$ compared with
Model~${\cal B}$ and the saturation value is larger for Model~${\cal U}$.
The inset of \Fig{pcomp} compares the results for Model~${\cal U}$
(with $\kf/k_1=15$) with Models ${\cal U}$10 and ${\cal U}$5 for $\kf/k_1=10$
and 5, respectively.
Note that NEMPI is quite weak for $\kf/k_1=5$, which is
consistent with the DNS presented here.

\begin{figure}\begin{center}
\includegraphics[width=\columnwidth]{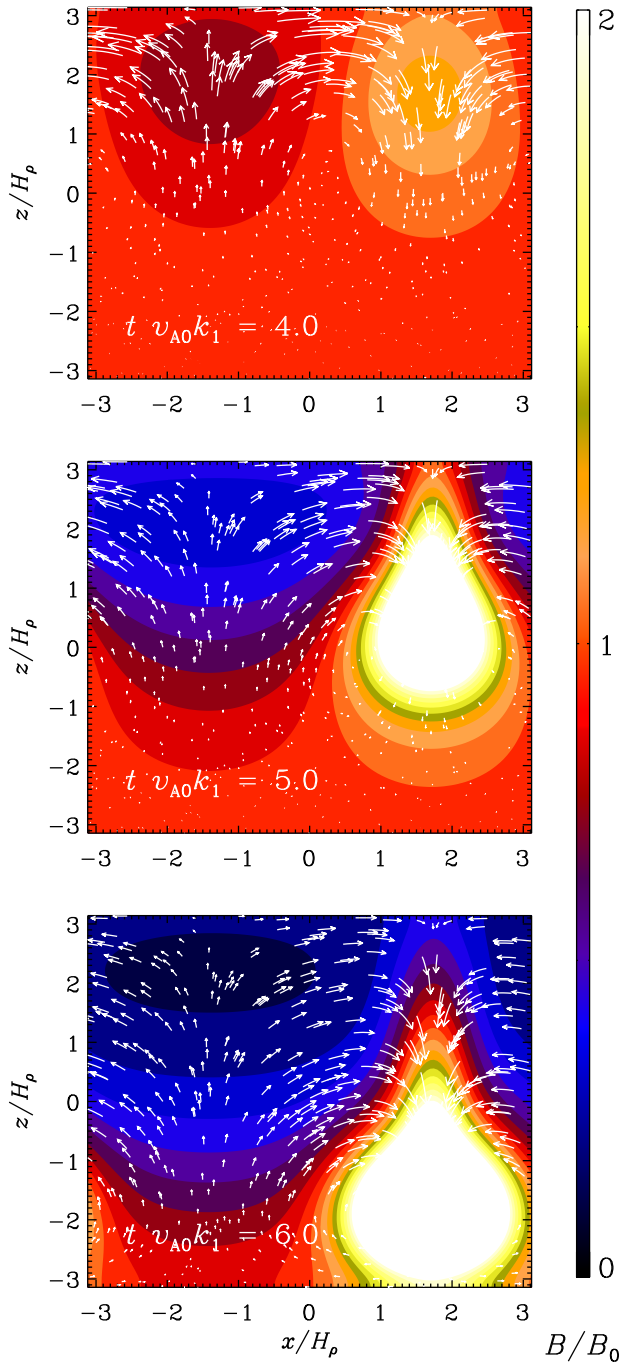}
\end{center}\caption[]{
Mean magnetic field in the $y$ direction (color coded)
together with velocity vectors in the $xz$ plane for Model~${\cal U}$02.
Note the spontaneous formation of flux structures.
}\label{pbcomp}\end{figure}

Visualizations of the mean magnetic field as well as the mean velocity
are shown in \Fig{pbcomp} for three different times near saturation
for Model~${\cal U}$02.
Here, we use a weaker imposed field, $B_0=0.002\rho^{1/2}\cs$, corresponding to
$B_0/\Beqz=0.02$, so that NEMPI starts closer to the surface.
For $B_0/\Beqz=0.05$, NEMPI starts in the middle of the domain, leaving
less space before the structures have reached the bottom of the domain.
Note the converging flows toward the magnetic structures, with the largest
velocities occurring in the upper layers where the density is smallest.

\begin{table}[h!]\caption{
Summary of parameters entering models U and B.
}\vspace{12pt}\centerline{\begin{tabular}{lrclccl}
Model&$\tilde\kf$&$\tilde\etatz$&$\tilde\Beqz$&$\beta_0$&$\Rm$ & comment\\
\hline
${\cal U}$  &15&0.0072&0.11 &0.05&  36   &$\Beq=\Beqz  \,e^{-z/2H_\rho}$ \\
${\cal U}$10&10&0.0048&0.11 &0.05&  36   &$\Beq=\Beqz  \,e^{-z/2H_\rho}$ \\
${\cal U}$5 & 5&0.0024&0.11 &0.05&  36   &$\Beq=\Beqz  \,e^{-z/2H_\rho}$ \\
${\cal U}$02&15&0.0072&0.11 &0.02&  36   &$\Beq=\Beqz  \,e^{-z/2H_\rho}$ \\
${\cal B}$  &15&0.0024&0.036&0.05&2.5--60&$\etat=\etatz\,e^{z/2H_\rho}$ \\
${\cal B}$0 &15&0.0024&0.036&0.05&2.5--60&$\gamma=0$ assumed \\
\label{Tsummary}\end{tabular}
\footnotetext{Tildes indicate nondimensional quantities: $\tilde\kf=\kf/k_1$,
$\tilde\etatz=\etatz k_1/\cs$, $\tilde\Beqz=\Beqz/\rho_0^{1/2}\cs$,
while $\beta_0=B_0/\Beqz$.}
}\end{table}

\subsubsection{Comments on the shape of mean-field structures}

The descending structures found in the present mean-field calculations are
qualitatively similar to those of BKR who considered a polytropic layer.
In both cases the structures sink and become wider.
This is quite different from the behavior of individual turbulent eddies
and flux tubes that one tends to monitor in DNS.
Clearly, individual magnetic structures experience magnetic buoyancy,
where the vertical motion is the result of a balance of magnetic buoyancy
and downward advection by the ambient flow \citep[see Fig.~10 of][]{BJNRST}.
The mean-field model cannot describe individual (small-scale) structures,
although the net effect of the vertical motion of individual
structures results in a turbulent pumping velocity $\ggamma$ of mean-field
structures.

While turbulent downward pumping has been seen in numerous DNS, the
negative effective magnetic pressure instability is a new effect that
has been seen so far only in the DNS of
forced turbulence in BKKMR; see also \cite{KBKMR}.
Amazingly, those DNS show very similar structures resembling that of a
descending ``potato sack'' (see Fig.~1 of BKKMR).
They sink because the negative effective magnetic pressure is compensated
by increased gas pressure, which in turn leads to larger density, so they
become heavier than the surroundings.
However, these turbulent magnetic structures are only poorly associated
with material motion; see the flow vectors in \Fig{pbcomp}.
Therefore, a change of their volume is not governed by mass conservation.
In particular, these structures do not become narrower during their descent
as individual blobs do in a strongly stratified layer.

We should point out that these mean-field structures do not always
initiate at the top of the layer.
The initiation height depends on the value of $z$ where $\Peff$
reaches a minimum.
Larger values of $B_0$ tend to move this location downward;
see \Fig{pxyaver_compb}.
A more detailed exploration of this is given by \cite{KBKR}.

\subsubsection{Comparison with the Parker instability}

NEMPI can be understood as a generalization of the Parker instability.
This becomes evident when considering the stability criterion of NEMPI (RK07):
\EQ
(H_{\rho} - H_{B}) \left.{\dd\Peff\over\dd\beta^2}\right|_{\beta_0} > 0,
\label{crit}
\EN
where $\dd\Peff/\dd\beta^2=\half(1-\qp-\dd\qp/\dd\ln\beta^2)$ and
$H_{B}$ is the characteristic
spatial scale of the mean magnetic field variations.
However, unlike the Parker instability,
NEMPI can be excited even in a uniform mean magnetic field
($H_{B} \to \infty$).
The source of free energy of this instability is provided by the
small-scale turbulent fluctuations.
In contrast, the free energy in the Parker's magnetic buoyancy
instability \citep{Par66} or in the interchange instability in plasma
\citep{Tse60} is drawn from the gravitational field.
In the absence of turbulence ($\qp=0$),
condition~(\ref{crit}) coincides with the criterion for the Parker's
magnetic buoyancy instability ($H_{\rho} > H_{B}$).

\section{Conclusions}
\label{Concl}

Our DNS have shown that for an isothermal atmosphere
with strong density stratification the total turbulent pressure is
decreased due to the generation of magnetic fluctuations
by the tangling of an imposed horizontal mean magnetic field
by the velocity fluctuations. This phenomenon strongly
affects the mean Lorentz force so that the effective mean
magnetic pressure becomes negative.
For our numerical model
with approximately uniform turbulent rms velocity, the
ratio of imposed to equipartition field strength changes with height,
because the density decreases with height, while the imposed field is constant.
This allows us to determine the full functional form of the effective
mean magnetic pressure as a function of normalized field strength for
a single run.

The form of the dependence of $\Peff(\meanB/\Beq)$ is similar to that
found for simulations under rather different conditions (with or
without stratification, with or without convection, etc).
This dependence is found to be similar to that obtained earlier using
both analytic theory (RK07) and direct numerical simulations (BKR),
and the results are robust when changing the strength of the imposed field.

In simulations where the turbulent velocity is nearly independent of
height, the reduction of magnetic fluctuations occurs
in the upper layers where the equipartition
field strength decreases with height (models of type U).
In models of type B, where the equipartition field strength is nearly constant
in height, the magnetic fluctuations are found to be slightly stronger
in the upper parts.

In view of astrophysical applications, it is encouraging that
$\qpz$ and $\betap$ seem to approach an asymptotic regime for $\Rm>60$.
While it remains important to confirm this result, a number of other
aspects need to be clarified.
Firstly, the issue of finite scale separation is important, i.e., the
larger the wavenumber of structures in the mean field relative to $\kf$,
the less efficient the negative effective magnetic pressure will be.
This needs to be quantified.
For example in the work of \cite{BKKMR11}, where we had a scale separation
ratio of 1:15, magnetic structures were best seen after averaging along the
direction of the mean field.
On the other hand, with a scale separation ratio of 1:30, structures
where quite pronounced already without averaging \citep{KBKMR}.
In the Sun, the scale separation ratio is very large in the horizontal
direction, but in the vertical direction the system is extremely
inhomogeneous and the vertical pressure scale height increases rapidly
with depth.
The significance of such effects on the negative effective magnetic
pressure effect remains still quite unclear.

Another aspect concerns the limitations imposed by the use of an
isothermal equation of state.
In the context of regular (non-turbulent) magnetic buoyancy,
the system is known to be more unstable to the buoyancy instability
when the fluctuations evolve isothermally \citep{Ach78,HW95,MC03}.
However, in the context of the negative effective magnetic pressure
instability it is not clear in which direction this effect would work.
The only simulations where the equation for specific entropy
was taken into account are the simulations of \cite{Kapy11},
who also considered an unstably stratified atmosphere.
In their case, the negative effective magnetic pressure was
found to be greatly enhanced (deeper minimum of $\Peff$ and
larger values of $\Bmin$).
Again, this is a subject that deserves serious attention.

The fact that the values of $\qpz$ and $\betap$ appear to be converged
in the range $60<\Rm<600$ is significant, because this is also the regime
in which small-scale dynamo action occurs.
Small-scale dynamo action suppresses the magnetic pressure effect,
which is the reason for the drop of $\qpz$ between $\Rm$ of 40 and 60,
but for larger $\Rm$, the values of $\qpz$ seem roughly unchanged.

In the present paper we have discussed applications mainly to the Sun.
However, any hydromagnetic turbulence with strong stratification
and large plasma beta should be subject to the negative effective magnetic pressure phenomenon.
Another relevant example might be accretion disks.
Quasi-periodic oscillations and other light curve variations from
accretion disks have long been suspected to be caused by some kind
of structures in these disks.
Hydrodynamic vortices would be one possibility \citep{Abramo}, which
could constitute long-lived structures \citep{Barge,Johansen,Lyra}.
However, in view of the present results, structures caused by the negative
effective magnetic pressure instability might indeed be another candidate.

\acknowledgments

We acknowledge the NORDITA dynamo programs of 2009 and 2011 for
providing a stimulating scientific atmosphere.
We acknowledge the allocation of computing resources provided by the
Swedish National Allocations Committee at the Center for
Parallel Computers at the Royal Institute of Technology in
Stockholm and the National Supercomputer Centers in Link\"oping
as well as the Norwegian National Allocations Committee at the
Bergen Center for Computational Science.
This work was supported in part by
the European Research Council under the AstroDyn Research Project No.\ 227952
and the Swedish Research Council Grant No.\ 621-2007-4064.
NK and IR thank NORDITA for hospitality and support during their visits.

\appendix

\section{Derivation of Equation (21)}
\label{SOCA}

We use the quasi-linear approach or second order correlation
approximation applied to a random flow with small magnetic
and fluid Reynolds numbers \citep[e.g.][]{Mof78,KR80,RKS12}.
We eliminate the pressure term
from the equation for the velocity fluctuations ${\bm u}$ by
calculating $\nab {\bm \times} (\nab {\bm \times} {\bm u})$,
rewrite the obtained equation and the induction
equation for the magnetic fluctuations ${\bm b}$
in Fourier space, apply the two-scale approach \citep{RS75}, and
neglect nonlinear terms, but retain molecular dissipative terms.
This allows us to get the following equation for
$\Delta\overline\Pi_{ij}^{f}$ from \Eq{stress0} in Fourier space:
\EQ
\Delta\overline\Pi_{ij}^{f}({\bm k}, \omega)= - \left[\hat L
\left(1+{G_\eta^\ast(k, \omega) \over G_\nu(k, \omega)}\right)
+ \hat L^\ast \right]\overline\Pi_{ij}^{f,0}({\bm k}, \omega),
\label{AP1}
\EN
where $G_\nu(k, \omega) = (\nu k^2 + i \omega)^{-1}$,
$G_\eta(k, \omega) = (\eta k^2 + i \omega)^{-1}$, $\hat L=G_\nu G_\eta
k_m k_n \meanB_m\meanB_n$, and the background turbulence (with a zero
mean magnetic field) is given by
\begin{eqnarray}
\overline\Pi_{ij}^{f,0}({\bm k}, \omega) =  {E(k) \, \Phi(\omega)
\over 8 \pi \, k^{2}}\, \Big(\delta_{ij} - {k_i\,k_j\over k^2}
\Big) \overline{{\bm u}^2_0}.
 \label{AP2}
\end{eqnarray}
Here $E(k)$ is the energy spectrum, e.g., a power-law
spectrum, $E(k) = (q-1) (k/\kf)^{-q} \, \kf^{-1}$ with exponent $1<q<3$
for the wavenumbers $\kf \leq k \leq \kd$, $\kf$ and $\kd$ are the forcing
and dissipation wavenumbers,
and we neglected for simplicity the anisotropy terms in Eq.~(\ref{AP2})
which are proportional to $\lambda_i$ and $\lambda_i \lambda_j$,
where $\lambda_i$ is a vector characterizing the anisotropy
(see Appendix~\ref{NonlinearCoefficients}).
We have taken into account that for small magnetic
and hydrodynamic Reynolds numbers the small-scale dynamo
is not excited, so that the background turbulence
contains only the velocity fluctuations.
We assume the frequency function $\Phi(\omega)$ to be a Lorentzian:
$\Phi(\omega)=1 / [\pi \tau_c \,(\omega^2 + \tau_c^{-2})]$.
This model for the frequency function
corresponds to the correlation function
$\langle u_i(t) u_j(t+\tau) \rangle \propto
\exp (-\tau /\tau_c)$.
After integration over $\omega$ and all angles in
${\bm k}$ space, and using Eq.~(\ref{PifB0}),
we arrive at the following equations for $\qpz$ and $\qsz$:
\begin{eqnarray}
\qpz = {\tau_c^2 \overline{{\bm u}^2_0}
\over 15} \int \Big(8 - {1 + \tau_c \nu k^2\over \tau_c \eta k^2}\Big)
\, {k^2 E(k) \over (1 + \tau_c \, \nu k^2) (1 + \tau_c \eta k^2)} \, dk,
 \label{AP3}\\
\qsz = {\tau_c^2 \overline{{\bm u}^2_0}
\over 15} \int \Big(2 + {1 + \tau_c \nu k^2\over \tau_c \eta k^2}\Big)
\, {k^2 E(k) \over (1 + \tau_c \, \nu k^2) (1 + \tau_c \eta k^2)} \, dk,
 \label{AP4}
\end{eqnarray}
where we take into account that $\beta \ll 1$.
In the derivation of Eqs.~(\ref{AP3}) and ~(\ref{AP4}) we used
the following integrals for the integration in $\omega$ space:
\begin{eqnarray}
\int G_\eta \,G_\nu \, G_\tau \,G^*_\tau \, d\omega = {\pi \tau_c^3
\over (1 + \tau_c \, \nu k^2) (1 + \tau_c \eta k^2)},
\quad
\int G_\eta \,G^*_\eta \, G_\tau \,G^*_\tau \, d\omega = {\pi \tau_c^2
\over \eta k^2 (1 + \tau_c \eta k^2)},
 \label{AP6}
\end{eqnarray}
where $G_\tau=(i \omega + \tau_c^{-1})^{-1}$.
We take into account that
for small magnetic and fluid Reynolds numbers $\tau_c \eta \kf^2 \gg 1$ and
$\tau_c \, \nu \kf^2 \gg 1$.
In this limit the coefficients $\qpz$ and $\qsz$,
after integration over $k$, are given by:
\begin{eqnarray}
\qpz = {C(q) \over 15} (8 - \Pm) \, \Rm \, \Rey,
\quad
\qsz = {C(q) \over 15} (2 + \Pm) \, \Rm \, \Rey,
\label{AP8}
\end{eqnarray}
where
\begin{eqnarray}
C(q) = \int_{\kf}^{\kd} E(k) \, \left({k\over \kf}\right)^{-2} dk
= \left({q-1\over q+1}\right) \,
\left[{1 - (\kf/\kd)^{q+1} \over 1 - (\kf/\kd)^{q-1}}\right],
 \label{AP9}
\end{eqnarray}
where $\Pm=\Rm / \Rey$.

\section{Theoretical $\meanB$ dependence of $\qp$, $\qs$, and $\qe$}
\label{NonlinearCoefficients}

In the following we summarize theoretical results for the $\meanB$
dependence of the coefficients $\qp$, $\qs$, and $\qe$ that enter in
\Eq{PifB0}.
These results were obtained for large magnetic and fluid Reynolds numbers
using the $\tau$ relaxation approach.
We recall that the coefficient $\qp$ represents the isotropic
turbulence contribution to the mean magnetic pressure, and
$\qg$ is the anisotropic turbulence contribution to the mean
magnetic pressure, while the coefficient $\qs$ is
the turbulence contribution to the mean magnetic tension.
We focus here on the case of anisotropic density-stratified background
turbulence.
Expressions for the isotropic case were given by
RK07 and are summarized in BKR.
Following BKR, we define $\beta \equiv \meanB/\Beq$.
We consider a plasma with a gas pressure that is much larger
than the magnetic pressure, and the total pressure
is always positive.

We define the scale of the energy-carrying eddies as $\ellf\approx\kf^{-1}$.
Due to density stratification, new terms emerge that are proportional
to $\ellf^2/H_\rho^2$.
These terms were absent in BKR, but otherwise the following formulae
are identical.
We also define the parameter
$\epsilon=\langle {\bm b}_0^2 \rangle/\langle {\bm u}_0^2 \rangle$, which
takes into account the contributions caused by the small-scale dynamo
(see RK07, where it was assumed for simplicity that the range of scales
of magnetic fluctuations generated by the small-scale dynamo coincides
with that of the velocity fluctuations).
\Tab{ModelsU} suggests $\epsilon=\betarms^2\approx0.3$.

For very weak mean magnetic fields, $4\beta \ll \Rm^{-1/4}$,
the values of $\qp$, $\qs$, and $\qg$ are approximately constant
and given by
\begin{equation}
q_{p}(\beta) = {4 \over 45} \, \big(1 + 9\ln \Rm\big) \, (1-\epsilon)
+ {16 \, \ellf^2 \over 9 \, H_\rho^2},\quad
q_{s}(\beta) = {1 \over 15} \, \big(1 + 8\ln \Rm\big) \, (1-\epsilon), \quad
\qg(\beta) = - {8 \, \ellf^2 \over 3 \, H_\rho^2};
\end{equation}
for $ \Rm^{-1/4} \ll 4\beta \ll 1$ we have
\begin{equation}
q_{p}(\beta) = {16 \over 25} \, [1 + 5|\ln (4 \beta)| + 32
\, \beta^{2}] \, (1-\epsilon) + {16 \, \ellf^2 \over 9 \, H_\rho^2}
\Big[1 - {16 \beta^{2} \over 5} \Big],
\end{equation}
\begin{equation}
q_{s}(\beta) = {32 \over 15} \, \biggl[|\ln (4 \beta)| +
{1 \over 30} + 12  \beta^{2} \biggr] \, (1-\epsilon),
\quad
q_{g}(\beta) =  - {8 \, \ellf^2 \over 3 \, H_\rho^2} \,
\Big[1 - {16 \, \beta^{2} \over 5} \Big] ;
\end{equation}
and for strong fields, $4\beta \gg 1 $, we have
\begin{equation}
q_{p}(\beta) = {1 \over 6\beta^2} \Big(1-\epsilon + {3 \, \ellf^2 \over \, H_\rho^2} \Big),
\quad q_{s}(\beta) = {\pi \over 48\beta^3}\, (1-\epsilon),
\quad q_{g}(\beta) = - {3 \, \ellf^2 \over 4 \, H_\rho^2 \, \beta^2}.
\end{equation}
Here we have taken into account that the anisotropic contributions to
the nonlinear functions $q_{p}(\beta)$ and $q_{g}(\beta)$ for
density-stratified background turbulence are given by
\begin{equation}
q_{g}(\beta) = - {3 \over 2} \, q_{p}(\beta)
= - {8 \, \ellf^2 \over 3 \, H_\rho^2} \,
\Big[64 \beta^4 - 4 \beta^2 + {1 \over 3} + {1 \over 4 \beta^2}
- 2^{9} \, \ln \Big(1 + {1 \over 8\beta^2} \Big) -
{\arctan (\sqrt{8} \beta) \over 8 \sqrt{2} \, \beta^3}  \Big] .
\label{anisotr}
\end{equation}
For the derivation of Eq.~(\ref{anisotr}) we used Eqs.~(A10)--(A11)
given by RK07 with the following model of
the density-stratified background turbulence written in the Fourier space:
\begin{equation}
\langle u_i({\bm k}) \, u_j(-{\bm k}) \rangle
={\langle {\bm u}_0^2 \rangle \, E(k) \over 8 \pi \, k^2 \,
(k^2 + \lambda^2)} \Big[\delta_{ij}\, (k^2 + \lambda^2) - k_i \, k_j
- \lambda_i \, \lambda_j
+ \ii \, \big(\lambda_i \, k_j - \lambda_j \, k_i\big) \Big] ,
\label{model}
\end{equation}
where the velocity field satisfies the continuity equation in
the anelastic approximation ${\rm div} \, {\bm u} = u_i \,
\lambda_i$, $\, \lambda_i = - \nabla_i \meanrho / \meanrho$,
the energy spectrum function is $E(k) = (2/3)
\, \kf^{-1} \, (k / k_{f})^{-5/3}$ for $\kf < k < \kf \, {\rm Re}^{3/4}$.

\section{Comments on the test-field method}
\label{TFM}

In the test-field method one uses a set of different
test fields to determine all relevant components of the $\alpha$ and
turbulent diffusivity tensors.
Furthermore, for finite scale separation ratios in space and
time one also needs to represent all relevant wavenumbers and frequencies.
The knowledge of all higher wavenumbers and frequencies allows one to compute
the integral kernels that describe the nonlocality of turbulent transport; see
\cite{BRS08} for nonlocality in space and \cite{HB09} for nonlocality in time.
The multitude of test fields does allow one to compute also those parts
of the $\alpha$ and turbulent diffusivity tensors that do not enter in
the particular problem at hand, but also those parts
that enter under any other circumstances.
An example is the evolution of a passive vector field where the same
mean-field theory applies \citep{TB08}.

Furthermore, given that we use the quasi-kinematic test-field method, we need
to address the work of \cite{CHP10}, who point out that
this method fails if there is
hydromagnetic background turbulence originating, for example, from
small-scale dynamo action.
In such a case a fully nonlinear test-field method must be employed
\citep[see][for details and implementation]{RB10}.
However, it is worth noting that even in cases where small-scale
dynamo action was expected, such as those of \cite{BRRS08} where values
of $\Rm$ up to 600 were considered, the quasi-kinematic test-field method
was still found to yield valid and self-consistent results,
as was demonstrated by comparing the growth rate expected from the
obtained coefficients of $\alpha_{ij}$ and $\eta_{ij}$.
This growth rate was confirmed to be compatible with zero in the
steady state.
Finally, as shown in \cite{RB10}, the quasi-kinematic method is
valid if magnetic fluctuations result solely from an imposed field.
In particular, the quasi-kinematic test-field method works even in
cases in which magnetic fluctuations are caused by a magnetic buoyancy
instability \citep{Chatterjee}.

\section{Comments on mean-field buoyancy}
\label{MeanFieldBuoyancy}

The work of \cite{KP93} is of interest in the present context, because
it predicts the upward pumping of mean magnetic field.
Here we discuss various aspects of this work.
\cite{KP93} assumed that:
(i) the gradient of the mean density is zero,
(ii) the background turbulence is homogeneous, and
(iii) the fluctuations of pressure, density and temperature are adiabatic.
We also note that their analysis is restricted to low Mach number flows,
although this is not critical for our present discussion.
Since the gradient of the mean density is zero, the hydrostatic
equilibrium, $\nab p = \rho {\bm g}$, exists only if the
gradient of the mean temperature is not zero. This implies that
the turbulent heat flux is not zero and temperature
fluctuations are generated by the tangling of this mean
temperature gradient by the velocity fluctuations.
Therefore, the key assumption made in \cite{KP93}
that fluctuations of pressure, density and
temperature are adiabatic, is problematic and the equation for
the evolution of entropy fluctuations should be taken into
account. This implies furthermore that the temperature fluctuations
in Eq.~(2.5) of their paper cannot be neglected.
We avoid this here by considering flows with a non-zero
mean density gradient and
turbulence simulations that have strong density stratification.



\begin{thebibliography}{}

\bibitem[Abramowicz et al.(1992)]{Abramo}
Abramowicz, M. A., Lanza, A., Spiegel, E. A., \& Szuszkiewicz, E.\ynat{1992}{356}{41}

\bibitem[Acheson(1978)]{Ach78}
Acheson, D. J.\yptrsa{1978}{289}{459}

\bibitem[Barge \& Sommeria(1995)]{Barge}
Barge, P., \& Sommeria, J.\yana{1995}{295}{L1}

\bibitem[Brandenburg(2011)]{B11}
Brandenburg, A.\yapj{2011}{741}{92}

\bibitem[Brandenburg \& Dobler(2002)]{BD02}
Brandenburg, A., \& Dobler, W.\yjour{2002}{Comp. Phys. Comm.}{147}{471}

\bibitem[Brandenburg \& Subramanian(2005a)]{BS05}
Brandenburg, A., \& Subramanian, K.\ 2005a, \physrep, 417, 1

\bibitem[Brandenburg \& Subramanian(2005b)]{BS05b}
Brandenburg, A., \& Subramanian, K.\yana{2005b}{439}{835}

\bibitem[Brandenburg \& Subramanian(2007)]{BS07}
Brandenburg, A., \& Subramanian, K.\yan{2007}{328}{507}

\bibitem[Brandenburg et al.(1996)]{BJNRST}
Brandenburg, A., Jennings, R. L., Nordlund, \AA.,
Rieutord, M., Stein, R. F., Tuominen, I.\yjfm{1996}{306}{325}

\bibitem[Brandenburg et al.(2004)]{BKM04}
Brandenburg, A., K\"apyl\"a, P., \& Mohammed, A.\ypf{2004}{16}{1020}

\bibitem[Brandenburg et al.(2011)]{BKKMR11}
Brandenburg, A., Kemel, K., Kleeorin, N., Mitra, D., \& Rogachevskii, I.\yapjl{2011}{740}{L50}

\bibitem[Brandenburg et al.(2010)]{BKR10}
Brandenburg, A., Kleeorin, N., \& Rogachevskii, I.\yan{2010}{331}{5} (BKR)

\bibitem[Brandenburg et al.(2008a)]{BRRK08}
Brandenburg, A., R\"adler, K.-H., Rheinhardt, M., \& K\"apyl\"a, P. J.\yapj{2008a}{676}{740}

\bibitem[Brandenburg et al.(2008b)]{BRRS08}
Brandenburg, A., R\"adler, K.-H., Rheinhardt, M., \& Subramanian, K.\yapjl{2008b}{687}{L49}

\bibitem[Brandenburg et al.(2008c)]{BRS08}
Brandenburg, A., R\"adler, K.-H., \& Schrinner, M.\yana{2008c}{482}{739}

\bibitem[Cattaneo \& Hughes(1988)]{Catt88}
Cattaneo, F., \& Hughes, D. W.\yjfm{1988}{196}{323}

\bibitem[Cattaneo et al.(1991)]{Cat91}
Cattaneo, F., Brummell, N. H., Toomre, J., Malagoli, A., and
Hurlburt, N. E.\yapj{1991}{370}{282}

\bibitem[Chatterjee et al.(2011)]{Chatterjee}
Chatterjee, P., Mitra, D., Rheinhardt, \& M. Brandenburg, A.\yana{2011}{534}{A46}

\bibitem[Courvoisier et al.(2010)]{CHP10}
Courvoisier A., Hughes D. W., \& Proctor M. R. E.\yprs{2010}{466}{583}

\bibitem[Dobler et al.(2002)]{DSB02}
Dobler, W., Shukurov, A., \& Brandenburg, A.\ypre{2002}{65}{036311}

\bibitem[Gilman(1970a)]{Gil70a}
Gilman P.A.\yapj{1970a}{162}{1019}

\bibitem[Gilman(1970b)]{Gil70b}
Gilman P.A.\yana{1970b}{286}{305}

\bibitem[Hood et al.(2009)]{HAGM09}
Hood, A. W., Archontis, V., Galsgaard, K., \& Moreno-Insertis, F.\yana{2009}{503}{999}

\bibitem[Hubbard \& Brandenburg(2009)]{HB09}
Hubbard, A., \& Brandenburg, A.\yapj{2009}{706}{712}

\bibitem[Hughes(2007)]{Hugh07}
Hughes, D. W. 2007, in The Solar Tachocline, ed. D. W. Hughes,
R. Rosner, \& N. O. Weiss (Cambridge: Cambridge Univ. Press), 275

\bibitem[Hughes \& Proctor(1988)]{HP88}
Hughes, D. W., \& Proctor, M. R. E.\yanf{1988}{20}{187}

\bibitem[Hughes \& Weiss(1995)]{HW95}
Hughes D. W., \& Weiss, N. O.\yjfm{1995}{301}{383}

\bibitem[Iskakov et al.(2007)]{Iska}
Iskakov, A. B., Schekochihin, A. A., Cowley, S. C., McWilliams, J. C., Proctor, M. R. E.\yprl{2007}{98}{208501}

\bibitem[Isobe et al.(2005)]{Isobe05}
Isobe, H., Miyagoshi, T., Shibata, K., \& Yokoyama, T.\ynat{2005}{434}{478}

\bibitem[Johansen et al.(2004)]{Johansen}
Johansen, A., Andersen, A. C., \& Brandenburg, A.\yana{2004}{417}{361}

\bibitem[K\"apyl\"a et al.(2012)]{Kapy11}
K\"apyl\"a, P. J., Brandenburg, A., Kleeorin, N., Mantere, M. J., \& Rogachevskii, I.\ymn{2012}{submitted}{arXiv:1104.4541}

\bibitem[Kemel et al.(2012a)]{KBKMR}
Kemel, K., Brandenburg, A., Kleeorin, N., Mitra, D., \&
Rogachevskii, I.\ysph{2012a}{in press}{arXiv:1112.0279}

\bibitem[Kemel et al.(2012b)]{KBKR}
Kemel, K., Brandenburg, A., Kleeorin, N., \& Rogachevskii, I.\yan{2012b}{333}{95}

\bibitem[Kersal\'{e} et al.(2007)]{Kers07}
Kersal\'{e}, E., Hughes, D. W., \& Tobias, S. M.\yapjl{2007}{663}{L113}

\bibitem[Kitchatinov \& Mazur(2000)]{KM00}
Kitchatinov, L.L., \& Mazur, M.V.\ysph{2000}{191}{325}

\bibitem[Kitchatinov \& Pipin(1993)]{KP93}
Kitchatinov, L. L., \& Pipin, V. V.\yana{1993}{274}{647}

\bibitem[Kitchatinov et al.(1994)]{Kitch94}
Kichatinov, L. L., R\"{u}diger, G., \& Pipin V. V.\yan{1994}{315}{157}

\bibitem[Kitiashvili et al.(2010)]{KKWM10}
Kitiashvili, I. N., Kosovichev, A. G., Wray, A. A., \& Mansour, N. N.\yapj{2010}{719}{307}

\bibitem[Kleeorin et al.(1993)]{KMR93}
Kleeorin, N., Mond, M., \& Rogachevskii, I.\ypfb{1993}{5}{4128}

\bibitem[Kleeorin et al.(1996)]{KMR96}
Kleeorin, N., Mond, M., \& Rogachevskii, I.\yana{1996}{307}{293}

\bibitem[Kleeorin \& Rogachevskii(1994)]{KR94}
Kleeorin, N., \& Rogachevskii, I.\ypre{1994}{50}{2716}

\bibitem[Kleeorin et al.(1989)]{KRR89}
Kleeorin, N.I., Rogachevskii, I.V., \& Ruzmaikin, A.A.\ysovl{1989}{15}{274}

\bibitem[Kleeorin et al.(1990)]{KRR90}
Kleeorin, N.I., Rogachevskii, I.V., \& Ruzmaikin, A.A.\yjetp{1990}{70}{878}

\bibitem[Krause \& R\"adler(1980)]{KR80}
Krause, F., \& R\"adler, K.-H.\ybook{1980}
{Mean-field magneto\-hydro\-dy\-na\-mics and dynamo theory}
{Pergamon Press, Oxford}

\bibitem[Lyra et al.(2011)]{Lyra}
Lyra, W., \& Klahr, H.\yana{2011}{527}{A138}

\bibitem[MacGregor \& Cassinelli(2003)]{MC03}
MacGregor, K. B., \& Cassinelli, J. P.\yapj{2003}{586}{480}

\bibitem[Mart\'{\i}nez et al.(2008)]{Mart08}
Mart\'{\i}nez, J., Hansteen, V. \& Carlson, M.\yapj{2008}{679}{871}

\bibitem[Moffatt(1978)]{Mof78}
Moffatt, H.K.\ybook{1978}
{Magnetic field generation in electrically conducting fluids}
{Cambridge University Press, Cambridge}

\bibitem[Newcomb(1961)]{New61}
Newcomb, W. A.\ypf{1961}{4}{391}

\bibitem[Parker(1966)]{Par66}
Parker, E.N.\yapj{1966}{145}{811}

\bibitem[Parker(1979a)]{Par79a}
Parker, E.N.\yapj{1979a}{230}{905}

\bibitem[Parker(1979b)]{Par79b}
Parker, E. N.\ybook{1979b} {Cosmical magnetic fields}{Oxford University Press, New York}

\bibitem[Prandtl(1925)]{Pra25}
Prandtl, L.\yjour{1925}{Zeitschr.\ Angewandt.\ Math.\ Mech.}{5}{136}

\bibitem[R\"adler(1969)]{Rad69}
R\"adler, K.-H.\yjour{1969}{Geod. Geophys. Ver\"off., Reihe II}{13}{131}

\bibitem[Rempel et al.(2009)]{Remp09}
Rempel, M., Sch\"{u}ssler, M., \& Kn\"{o}lker, M.\yapj{2009}{691}{640}

\bibitem[Rheinhardt \& Brandenburg(2010)]{RB10}
Rheinhardt, M., \& Brandenburg, A.\yana{2010}{520}{A28}

\bibitem[Roberts \& Soward(1975)]{RS75}
Roberts, P. H., \& Soward, A. M.\yan{1975}{296}{49}

\bibitem[Rogachevskii \& Kleeorin(2006)]{RK06}
Rogachevskii, I., \& Kleeorin, N.\ygafd{2006}{100}{243}

\bibitem[Rogachevskii \& Kleeorin(2007)]{RK07}
Rogachevskii, I., \& Kleeorin, N.\ypre{2007}{76}{056307} (RK07)

\bibitem[R\"udiger(1980)]{Rue80}
R\"udiger, G.\ygafd{1980}{16}{239}

\bibitem[R\"udiger(1989)]{Rue89}
R\"udiger, G.\ybook{1989}{Differential rotation and stellar convection:
Sun and solar-type stars}{Gordon \& Breach, New York}

\bibitem[R\"udiger \& Hollerbach(2004)]{RH04}
R\"udiger, G., \& Hollerbach, R.\ybook{2004}{The magnetic universe}
{Wiley-VCH, Weinheim}

\bibitem[R\"udiger et al.(2012)]{RKS12}
R\"udiger, G., Kitchatinov, L. L.\& Schultz, M.\yan{2012}{333}{84}

\bibitem[Schrinner et al.(2005)]{Sch05}
Schrinner, M., R\"adler, K.-H., Schmitt, D., Rheinhardt, M., Christensen, U.\yan{2005}{326}{245}

\bibitem[Schrinner et al.(2007)]{Sch07}
Schrinner, M., R\"adler, K.-H., Schmitt, D., Rheinhardt, M., Christensen, U. R.\ygafd{2007}{101}{81}

\bibitem[Sch\"{u}ssler \& V\"{o}gler(2006)]{Schus06}
Sch\"{u}ssler, M., \& V\"{o}gler, A.\yapjl{2006}{641}{L73}

\bibitem[Solanki et al.(2006)]{Solanki}
Solanki, S. K., Inhester, B., \& Sch\"ussler, M.\yrpp{2006}{69}{563}

\bibitem[Stein \& Nordlund(2001)]{Stein01}
Stein, R. F. \& Nordlund, {\AA}.\yapj{2001}{546}{585}

\bibitem[Tao et al.(1998)]{Tao98}
Tao, L., Weiss, N. O., Brownjohn, D. P., \& Proctor, M. R. E.\yapj{1998}{496}{L39}

\bibitem[Taylor(1921)]{Tay21}
Taylor, G. I.\yjour{1921}{Proc. Lond. Math. Soc.}{20}{196}

\bibitem[Tilgner \& Brandenburg(2008)]{TB08}
Tilgner, A., \& Brandenburg, A. 2008, MNRAS, 391, 1477

\bibitem[Tobias \& Weiss(2007)]{Tob07}
Tobias, S. M., \& Weiss, N. O. 2007, in The Solar Tachocline, ed. D.W. Hughes,
R. Rosner, \& N. O. Weiss (Cambridge: Cambridge Univ. Press), 319

\bibitem[Tserkovnikov(1960)]{Tse60}
Tserkovnikov, Y. A. 1960, Sov. Phys. Dokl., 5, 87

\bibitem[Wissink et al.(2000)]{Wis00}
Wissink, J. G., Hughes, D. W., Matthews, P. C., \& Proctor, M. R. E.\ymn{2000}{318}{501}

\bibitem[Yousef et al.(2003)]{YBR03}
Yousef, T. A., Brandenburg, A., \& R\"udiger, G.\yana{2003}{411}{321}

\end{thebibliography}
\end{document}